\newif\ifcheckpagelimits
\checkpagelimitstrue
\checkpagelimitsfalse

\ifcheckpagelimits
 \documentclass[nofootinbib,prb,aps,twocolumn,showpacs,showkeys,%
 amsmath,amssymb,superscriptaddress,final,reprint,floatfix,longbibliography]{revtex4-1}
 \newcommand{\todo}[1]{}
\else
 \documentclass[prb,aps,twocolumn,showpacs,showkeys,%
 amsmath,amssymb,superscriptaddress,final,reprint,floatfix,longbibliography]{revtex4-1}
 \newcommand{\todo}[1]{{\pdfmargincomment[icon=Note,color=pink]{#1}}}
\fi

\usepackage{lineno}
  \usepackage{mathptmx}
\usepackage{subfigure}
\usepackage{dcolumn}
\usepackage{amsmath,amssymb}
\usepackage{bm}
\usepackage{color}
\usepackage{overpic}
\usepackage{latexsym}
\usepackage{epstopdf}
\usepackage{color}
\usepackage[english]{babel}
\usepackage{latexsym}
\usepackage{stmaryrd}

\usepackage{braket}

\definecolor{mygrey}{gray}{0.35}
\definecolor{myblue}{rgb}{0.2,0.2,0.8}
\definecolor{myzard}{cmyk}{0,0,0.05,0}
\definecolor{mywhite}{rgb}{1,1,1}
\definecolor{myred}{rgb}{1,0.,0.3}

\usepackage[colorlinks=true,citecolor=myblue,linkcolor=myred]{hyperref}
\DeclareMathAlphabet{\mathpzc}{OT1}{pzc}{m}{it}

 \def\ee{\mathord{\rm e}}
 
 \def\ii{\mathord{\rm i}}

\def\half{\textstyle\frac{1}{2}}

\renewcommand{\ii}{{\rm i}}
\renewcommand{\ee}{{\rm e}}
\def\beq{\begin{equation}}
\def\eeq{\end{equation}}

\def\barray{\begin{eqnarray}}
\def\earray{\end{eqnarray}}

\begin{document}

\title{Long-range Heisenberg  models in quasi-periodically driven crystals of  trapped ions }

\author{A. Bermudez}
\affiliation{Department of Physics, Swansea University, Singleton Park, Swansea SA2 8PP, United Kingdom}
\affiliation{Instituto de F\'{i}sica Fundamental, IFF-CSIC, Madrid E-28006, Spain}

\author{L. Tagliacozzo}
\affiliation{Department of Physics and SUPA, University of Strathclyde, Glasgow G4 0NG, United Kingdom}

\author{G. Sierra}
\affiliation{Instituto de F\'{i}sica Te\'{o}rica (IFT), UAM-CSIC, Madrid, Spain}

\author{P. Richerme}
\affiliation{Department of Physics, Indiana University, Bloomington, Indiana 47405, USA}

\begin{abstract}
We introduce  a theoretical scheme for the analog quantum simulation of   long-range XYZ models   using current trapped-ion technology. In order to achieve fully-tunable  Heisenberg-type interactions, our proposal requires a state-dependent dipole force along a single  vibrational axis, together with a combination of standard resonant and detuned  carrier drivings. We discuss how  this quantum simulator could explore  the effect of  long-range interactions on the phase diagram  by combining an adiabatic protocol with the quasi-periodic drivings, and test the validity of our scheme numerically. At the isotropic Heisenberg point, we show that the long-range  Hamiltonian  can be mapped onto a non-linear sigma model with a topological term that is responsible for its low-energy properties, and we benchmark our predictions with Matrix-Product-State numerical simulations.

\end{abstract}

\ifcheckpagelimits\else
\maketitle
\fi

 \ifcheckpagelimits
\else

\section{Introduction}
The dream of  designing the  Hamiltonian   of an atomic  system   $H(t)$  to reproduce a relevant   model   of  condensed-matter or high-energy physics $H_{\rm target}$~\cite{feynman_qs} is already  an experimental reality.  Quantum-optical setups of neutral atoms~\cite{QS_cold_atoms} and trapped ions~\cite{QS_trapped_ions} have  become highly-controllable platforms to address the  quantum many-body problem from a different perspective,  opening   interesting prospects for the short term~\cite{qs_goals}. Often, the theoretical   goal  is  to describe the dynamics of these    systems by a combination of  unitaries ($\hbar=1$)
\beq
\label{eq:target_U}
U(t):=\mathpzc{T}\!\left\{\ee^{-\ii\int_0^t {\rm d}\tau H(\tau)}\right\}\to U_{\rm eff}(t):=U_0(t)\ee^{-\ii H_{\rm eff} t},
\eeq
 where  $U_0(t)$  depends on the
scheme  used to target  the desired  model $H_{\rm eff}=H_{\rm target}$. This can be accomplished by Trotterizing the time evolution~\cite{digital_QS}, or by using an always-on Hamiltonian, which leads to the notions of digital/analog quantum simulators (QS).
 For periodic Hamiltonians $H(t)=H(t+T)$, an exact identity  can be found $U(t)=U_{\rm eff}(t)$ by   Floquet theory~\cite{floquet}. This has led to the concept of {\it Floquet engineering} and has important  applications in ultracold atoms~\cite{eckardt_review}. To gain further tunability over $H_{\rm eff}$,  one may  use a collection of drivings with different periodicities that need not be commensurate. However, a rigorous generalisation of  Floquet theory to these  {\it quasi-periodic drivings} is still an open problem~\cite{verdeny}. One thus searches for  schemes  where  Eq.~\eqref{eq:target_U} is  achieved approximately $U(t)\approx U_{\rm eff}(t)$, but  with sufficient accuracy, and where the additional effects brought up by $U_0(t)$ are not in conflict with the    target model of the analog quantum simulator.

In  trapped ions~\cite{wineland_review}, an analog QS~\eqref{eq:target_U}  for  the quantum Ising model,  a paradigm  of   phase transitions~\cite{itf,sachdev_book}, has already been achieved~\cite{porras_spin_models_ions,Ising_exp_ions,varenna_proceedings}. Unfortunately, implementing  what is arguably the most important model of low-dimensional  magnetism~\cite{Heisenberg-review,Heisenberg,Heisenberg_exact_solution_bethe}, the so-called {\it XYZ model}~\cite{xyz_exact_solution}
\beq
\label{eq:XYZ}
H_{\rm XYZ}=\sum_{i>j} \left(J_{ij}^x\sigma_i^{x}\sigma_j^{x}+J_{ij}^y\sigma_i^{y}\sigma_j^{y}+J_{ij}^z\sigma_i^{z}\sigma_j^{z}\right),
\eeq
where $J_{ij}^\alpha$ and $\sigma_i^\alpha$ are  coupling strengths and Pauli matrices for $\alpha\in\{x,y,z\}$, has remained elusive for several  years. Although the digital QS of this model has already been achieved~\cite{digital_ions}, and a combination of digital and analog protocols proposed~\cite{digital_analog_QS}, it would be desirable to find  purely-analog QS that can be scaled to larger ion crystals without the need of quantum error correction to mitigate  Trotterization errors.

In this work, we present such a  scheme using quasi-periodic drivings. To implement a rotated XXZ model  $H_{\rm eff}=H_{\rm XXZ}$ obtained from~\eqref{eq:XYZ} after setting $J_{ij}^y=J_{ij}^z$~\cite{xxz_exact_solution},  our scheme only requires modifying the  driving fields  that produce an effective Ising model~\cite{Ising_exp_ions}. Moreover, we show that the  XYZ model  can  be achieved  $H_{\rm eff}=H_{\rm XYZ}$ by introducing additional  drivings. We test the validity of our proposal against numerical simulations, and present a detailed study of the effect of the long-range interactions that   arise naturally in ion traps, and break the  integrability of the model~\cite{Heisenberg_exact_solution_bethe,xxz_exact_solution,xyz_exact_solution}. 

This article is organized as follows. In Sec.~\ref{sec:scheme}, we describe the scheme based on quasi-periodic drivings that leads to an effective   XYZ model for a trapped-ion crystal. We analyze the suitability of this scheme with respect to experimentally-available tools in Sec.~\ref{sec:exp}, and test its validity by comparing the analytical predictions to numerical simulations in Sec.~\ref{sec:num}. In Sec.~\ref{sec:NLSM}, we derive an effective quantum field theory to describe the low-energy properties of the effective long-range  XYZ model in the SU(2) symmetric regime, and tests some of its predictions using numerical algorithms based on Matrix-Product-States. Finally, we present our conclusions in Sec.~\ref{sec:conc}. Details of the different derivations in these sections are given in the Appendixes.  

\section{ Quasi-periodically driven  ions}
\label{sec:scheme}  We consider a trapped-ion chain subjected to different  drivings   between two electronic states $\ket{{\uparrow}},\ket{{\downarrow}}$~\cite{wineland_review}. The bare dynamics  is described by
$
H_0=\sum_{i}\frac{\omega_0}{2}\sigma_i^z+\sum_{n,\alpha}\omega_{n,\alpha} a_{n,\alpha}^{\dagger}a_{n,\alpha}^{\phantom{\dagger}},
$
where $\omega_0$ is the transition frequency,  and $a_{n,\alpha}^{\dagger}, a_{n,\alpha}^{\phantom{\dagger}}$  the phonon creation-annihilation operators in a normal mode of frequency $\omega_{n,\alpha}$~\cite{normal_modes}. A   useful quasi-periodic driving 
 is a  dipole force  transversal to the chain, e.g. M$\o$lmer-S$\o$rensen (MS)~\cite{molmer_sorensen}   force 
\beq
\label{eq:MS_force}
H_{\rm MS}=\sum_{i,n}\mathcal{F}_{in}x_0\sigma_i^+\left(a_{n,x}^{\phantom{\dagger}}\ee^{\ii\phi_{\rm r}-\ii \omega_{\rm r} t}+a^\dagger_{n,x}\ee^{\ii\phi_{\rm b}-\ii \omega_{\rm b} t}\right)+{\rm H.c.},
\eeq
where we have introduced the  light forces $\mathcal{F}_{in}$, the zero-point motion $x_0$ along the $x$ axis, $\sigma_i^+=\ket{{\uparrow}_i}\bra{{\downarrow}_i}$, and the frequencies $\omega_{\rm r}=\omega_0-\omega_{n,x}+\delta_n, \omega_{\rm b}=\omega_0+\omega_{n,x}-\delta_n$, and phases $\phi_{\rm r}, \phi_{\rm b}$, of two laser beams (see below). Another useful periodic driving follows from a laser/microwave  coupled to the transition
\beq
\label{eq:carrier}
H_{\rm C,1}=\sum_i h_{0}\sigma_i^+\ee^{\ii\phi_{\rm d}-\ii \omega_{\rm d} t}+{\rm H.c.},
\eeq
where $h_0$ is half the Rabi frequency, and we have introduced the driving frequency $\omega_{\rm d}$, and phase $\phi_{\rm d}$.

As shown  in theory and experiments~\cite{porras_spin_models_ions,Ising_exp_ions},  the quasi-periodic Hamiltonian  $H(t)=H_0+H_{\rm MS}+H_{\rm C,1}$,  in the  regime
\beq
\label{eq:regime_ITF}
 h_0\ll \mathcal{F}_{in} x_0\ll\delta_n\ll\omega_{n,x},\hspace{2ex}\omega_{\rm d}=\omega_0,
\eeq
leads to a time-evolution of the form~\eqref{eq:target_U} with $U(t)\approx U_{\rm eff}(t)$ targeting  a  long-range quantum Ising model $H_{\rm eff}=H_{\rm QIM}$ 
\beq
\label{eq:Ising}
H_{\rm QIM}=\sum_{i>j} J_{ij}\sigma_i^{\phi_{\rm s}}\sigma_j^{\phi_{\rm s}}+h_0 \sum_i\sigma_i^{\phi_{\rm d}}.
\eeq
Here, we have introduced the spin-spin couplings $J_{ij}=-\sum_n\mathcal{F}_{in}^{\phantom{*}}\mathcal{F}_{jn}^*x_0^2/\delta_n+{\rm c.c.}$,  $\sigma_i^{\theta}=\sigma_i^+\ee^{\ii\theta}+{\rm H.c.}$, and the spin phase $\phi_{\rm s}=(\phi_{\rm r}+\phi_{\rm b})/2+\pi/2$ of the MS lasers. Provided that the phases  fulfil $\phi_{\rm s}=\phi_{\rm d}+\pi/2$, the engineered Hamiltonian~\eqref{eq:Ising} corresponds to a transverse-field Ising model~\cite{itf,sachdev_book}. We note that the additional unitary in~\eqref{eq:target_U} is simply $
U_0(t)=\ee^{-\ii t H_0}$, which does not compromise the  measurement in the QS, e.g. magnetisation and  spin correlations. 

As outlined  in~\cite{porras_spin_models_ions}, by combining three  dipole forces along each vibrational axis, one may exploit all  phonon branches to mediate a Heisenberg-type interaction~\eqref{eq:XYZ}. However, in addition to the technical overhead of combining all the required laser beams,  there are some fundamental  limitations. Since the axial trap frequency  must decrease with the number of ions,  the spin-spin interactions mediated by axial phonons become weaker as the crystal grows. This becomes especially troublesome as the  $J^z_{ij}\sigma_i^z\sigma_j^z$ interactions require a differential ac-Stark shift, and thus exclude  using  clock states~\cite{clock}, making the experiment less resilient to magnetic-field fluctuations.
 Moreover, the  distance dependence of  axial- and transverse-mediated interactions differs markedly~\cite{porras_spin_models_ions}, such that the important SU(2)-symmetric point $J_{ij}^x=J_{ij}^y=J_{ij}^z$ cannot be achieved. Although some of these problems may be circumvented with surface traps~\cite{qs_surface_traps},  it would be  desirable to implement Heisenberg-type models with current Paul/Penning designs~\cite{Ising_exp_ions,Ising_exp_penning}, which   requires   using a single  branch  of transverse phonons to mediate the  interactions. Our main result  is to present such scheme  by combining the MS force~\eqref{eq:MS_force}, such that clock states can  be used to encode the spin, 
 with a  carrier term~\eqref{eq:carrier} supplemented by two additional tones
\beq
\label{eq:3_tone}
H_{\rm C,3}=\sum_i\sum_{{\rm t}=1,2,3} h_{\rm t}\sigma_i^+\ee^{\ii\phi_{\rm d,{\rm t}}-\ii \omega_{\rm d,{\rm t}} t}+{\rm H.c.},
\eeq
symmetrically detuned with respect to the carrier transition (see  Table~\ref{tab_1}). We show that the quasi-periodically driven Hamiltonian
$
H(t)=H_0+H_{\rm MS}+H_{\rm C,3},
$ leads to $U(t)\approx U_{\rm eff}(t)$  described by the  XYZ model $H_{\rm eff}=H_{\rm XYZ}$~\eqref{eq:XYZ}, where     the spin-spin interactions
\beq
\label{eq:coupling_constants}
\begin{split}
J_{ij}^x=J_{ij}\cos^2\phi_{\rm s},\hspace{2ex}J_{ij}^{y/z}=\half J_{ij}\sin^2\phi_{\rm s}(1\mp\mathpzc{J}_1(\xi)),
\end{split}
\eeq
are expressed in terms of the $J_{ij}$ couplings introduced below Eq.~\eqref{eq:Ising}, and the first-order Bessel function $\mathpzc{J}_1(x)$. Varying the spin phase $\phi_{\rm s}$ allows for independent control  over  $J_{ij}^x$ and $J_{ij}^{y/z}$, while varying the drive strengths $h_2$ and $h_3$ controls the asymmetry between $J_{ij}^y$ and $J_{ij}^z$. Moreover, the additional unitary in~\eqref{eq:target_U} is  $
U_0(t)=\ee^{-\ii t H_0}\ee^{-\ii \sum_i h_0 (t+ \xi\sin(\Delta t)/\Delta)\sigma_i^x}$. 

\begin{table}
\centering
  \caption{{\bf Parameters of the three-tone   driving in Eq.~\eqref{eq:3_tone}}  }
\begin{tabular}{ c|  c c  c  }
\hline \hline
{\rm Tone}& ${\rm t}=1$ & ${\rm t}=2$ &${\rm t}=3$  \\
\hline
$\hspace{2ex}{\rm Frequency}\hspace{2ex}$& $\hspace{2.5ex}\omega_{\rm d,{\rm 1}}=\omega_0$ & $\hspace{2.5ex}\omega_{\rm d,{\rm 2}}=\omega_0+\Delta\hspace{2ex}$ &$\omega_{\rm d,{\rm 3}}=\omega_0-\Delta\hspace{2ex}$ \\
${\rm Strength}$& $h_{\rm 1}=h_0$ & $h_{\rm 2}=\half h_0\xi$ &$h_{\rm 3}=\half h_0\xi$ \\
${\rm Phase}$& $\phi_{\rm d,  1}=0$ & $\phi_{\rm d,  2}=0$ &$\phi_{\rm d,  3}=0$ \\
\hline
\hline
\end{tabular}
\label{tab_1}
\end{table}

We now address the crucial task of finding the parameter regime that substitutes Eq.~\eqref{eq:regime_ITF}, and gives rise to a XYZ model~\eqref{eq:XYZ} instead of the usual Ising model~\eqref{eq:Ising}. Previous results  found that by modifying the  strength of the driving~\eqref{eq:carrier},   one  either obtains a new Ising model with the phase of the driving   for  $ \mathcal{F}_{in} x_0\ll \delta_n\ll h_0$~\cite{bermudez_gate}, or an isotropic XY model for  milder drivings  $ {\rm max}\{ \mathcal{F}_{in} x_0, h_0\}\ll \delta_n$, and $J_{ij}\ll h_0$~\cite{xy_dynamics}. If instead of the resonant  driving~\eqref{eq:carrier}, a periodically-modulated one is considered, it is possible to engineer an anisotropic XY model~\cite{duality}. We also note that more generic periodic drivings with a site-dependent phase allow to control also the directionality of the XY interactions, and achieve effective spin Hamiltonians corresponding to quantum compass models~\cite{porras_compass}.

These results thus suggest that we should combine  resonant and off-resonant  drivings, as in Eq.~\eqref{eq:3_tone}, and explore the regime of large, but not too strong, driving strengths.

 By using the  Magnus expansion~\cite{magnus}, together with  techniques for periodically-modulated  systems~\cite{pat,eckardt_review} (see Appendix~\ref{app:A}), we find that the  regime to obtain a XYZ model~\eqref{eq:XYZ} is
\beq
\label{eq:regime_XYZ_1}
{\rm max}\{\mathcal{F}_{in} x_0, h_0,\Delta\}\ll\delta_n\ll\omega_{n,x},\hspace{2ex}\xi<\half,
\eeq
together with
\beq
\label{eq:regime_XYZ_2}
{\rm max}\{J_{ij}\}\ll 2h_0,\hspace{2ex}\Delta=4h_0.
\eeq
 Condition~\eqref{eq:regime_XYZ_1} is important to {\it (i)} avoid that the  drivings perturb  the laser-ion interaction  leading to the MS force~\eqref{eq:MS_force}, and  {\it (ii)} minimise residual spin-phonon terms   impeding a description of the spin dynamics by a periodically-modulated Ising model
\beq
\label{eq:time_dep_ising}
H_{\rm eff}(t)=\sum_{i>j}J_{ij}\sigma_i^{\phi_{\rm s}}(t)\sigma_j^{\phi_{\rm s}}(t), \hspace{2ex}\sigma_i^{\phi_{\rm s}}(t)=\hat{U}(t)\sigma_i^{\phi_{\rm s}}\hat{U}(t)^\dagger,
\eeq 
where $\hat{U}(t)=\ee^{-\ii \sum_i h_0 (t+ \xi\sin(\Delta t)/\Delta)\sigma_i^x}
$. Finally, condition~\eqref{eq:regime_XYZ_2} guarantees that {\it (iii)} this periodically-modulated Ising model leads to the  desired XYZ Hamiltonian~\eqref{eq:XYZ}.

\begin{figure*}
\centering
\includegraphics[width=2\columnwidth]{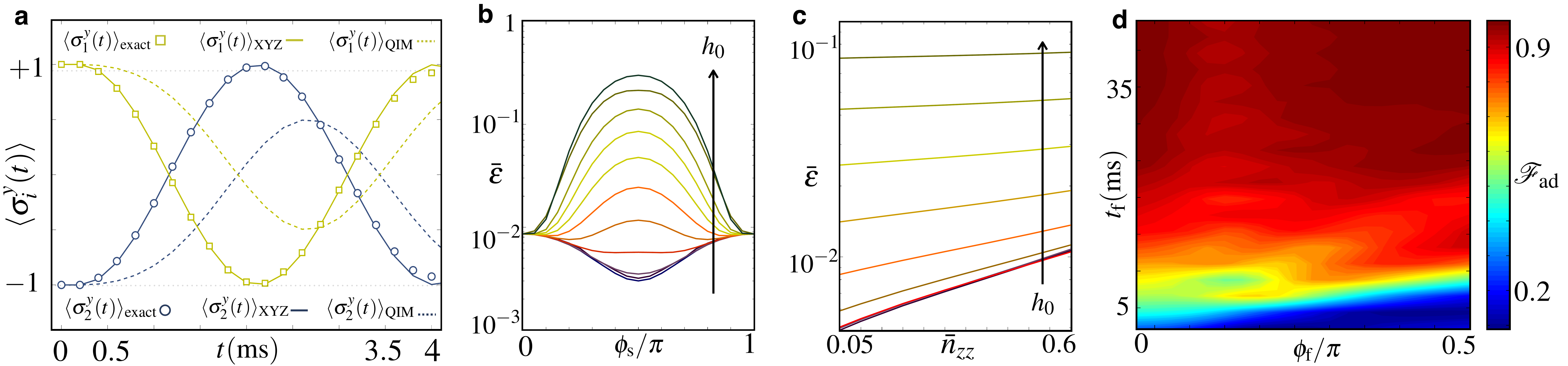}
\caption{ {\bf Numerical validation of the XYZ Heisenberg model:} {\bf (a)} Magnetization dynamics $\langle \sigma_i^y(t)\rangle_\beta$ for a two-ion chain evolving under  $\beta\in\{\rm exact, XYZ~\eqref{eq:XYZ},QIM~\eqref{eq:Ising}\}$. The  trapped-ion parameters for the MS~\eqref{eq:MS_force} are  $\delta/2\pi=500$kHz, $\Omega_{\rm L}/2\pi=0.9$MHz, and $\phi_{\rm s}=\pi/3$; whereas  for the modulated carrier~\eqref{eq:3_tone} in Table~\ref{tab_1},  $h_0/2\pi=2.5$kHz,  $\xi=0.09$ and  $\Delta/2\pi=10$kHz. We consider initial mean phonon numbers $\bar{n}_{zz}=0.05$, and $\bar{n}_{\rm com}=0.047$, and truncate the vibrational Hilbert space to 7 phonons per mode.
To ease the visualisation, we  perform a spin-echo refocusing pulse at the middle $U_{\rm s}=\ee^{\ii\frac{\pi}{2}\sum\sigma_i^y}$ for different evolution times  given by  multiples of $\pi/h_0=4\pi/\Delta$. {\bf (b,c)}   Quantum channel error as a function of {\bf (b)} the MS phase $\phi_{\rm s}$, and {\bf (c)} the mean number of phonons in the lowest mode, considering for the  same parameters, but setting $\Omega/2\pi=0.5$MHz, and varying $h_0/2\pi\in\{ 1.2,    1.8,    2.4,    3,    3.6,    4.2,    4.8,    5.3,    5.9\}$kHz, and the associated  detunings $\Delta=4h_0$.  The different lines correspond to the above values of $h_0$ increasing in the direction of the arrow. {\bf (d)} Adiabatic evolution of  $\ket{\psi_0}=\ket{-_y+_y-_y+_y-_y+_y}$ for $N=6$ ions, subjected to {\it (i)} an Ising linear ramp  $t\in[0 t_{\rm f}/2]$ of the staggered field $h_0(t)=h_0(0)-\delta h t$, with $h_0(0)=12J_{12}$, and $\delta h=-2h_0(0)/t_{\rm f}$, followed by {\it (ii)} a Heisenberg linear ramp $t\in[t_{\rm f}/2, t_{\rm f}]$ of the phase $\phi_{\rm s}(t)=\delta\phi t$, where $\delta \phi=2\phi_{\rm f}/t_{\rm f}$, as described in the text.  The adiabatic fidelity $\mathcal{F}_{\rm ad}$ is represented as a function of the total   ramp time $t_{\rm f}$, which is set to be an integer multiple of $2\pi/h_0$, and the final phase $\phi_{\rm f}$.}
\label{fig_num_validation}
\end{figure*}

\section{Experimental considerations for the scheme} 
\label{sec:exp}
 While our results are applicable to most ion species, and to Penning traps,  we focus on a chain of  $^{171}$Yb$^+$ ions  in a linear Paul trap biased to yield  transverse and axial trap frequencies of 5 MHz and 1 MHz, respectively,  and 2-3 $\mu$m ion spacing~\cite{xy_dynamics}. Within each Yb ion, the $^2$S$_{1/2}\ket{F=0,m_F=0}$ and $^2$S$_{1/2}\ket{F=1,m_F=0}$ hyperfine `clock' states, denoted $\ket{\uparrow}$ and $\ket{\downarrow}$ respectively,  encode the   effective spin-1/2 system \cite{YbDetection}.

The spin-spin interaction and external magnetic field  in Eq.~\eqref{eq:Ising} are routinely generated  by globally driving stimulated Raman transitions between the  spin states~\cite{Ising_exp_ions}. Two Raman beams are aligned with  their wavevector difference  along the transverse vibrational axis  ${\bf e}_{x}$ (see Appendix~\ref{app:A}). One is held at fixed frequency $\omega_{\rm L}$, while the other  contains multiple  frequencies $\omega_{\rm L}+\delta\omega_\ell$ imprinted by an acousto-optic modulator (AOM). The AOM is driven by an arbitrary waveform generator (AWG), allowing for full frequency, amplitude, and phase control over the components of the second Raman beam. For instance, simultaneous application of $\delta \omega_\ell \in \{\omega_{\rm r},\omega_{\rm b}\}$ with respective phases $\{\phi_{\rm r},\phi_{\rm b}\}$ will lead to a MS force~\eqref{eq:MS_force} that yields the first term of~\eqref{eq:Ising} if the parameters fulfil~\eqref{eq:regime_ITF}. Additionally,  applying $\delta\omega_\ell = \omega_0$ with phase $\phi_{\rm d}$ leads to~\eqref{eq:carrier} and  yields the second term of Eq.~\eqref{eq:Ising}. Typical  parameters are a carrier Rabi frequency \mbox{$\Omega_{\rm L}/2\pi\sim0.1$-$1$ MHz}, a Lamb-Dicke parameter $\eta=0.07$, and a MS detuning $\delta_n/2\pi\sim 100$-$500$ kHz, giving a maximum spin-spin coupling \mbox{$J_{\text{max}}/2\pi\sim 0.1$-$ 1$ kHz}. In the current scheme, choosing $h_0/2\pi\sim 1$-$ 10$ kHz  simultaneously satisfies the conditions in Eqs.~\eqref{eq:regime_XYZ_1} and~\eqref{eq:regime_XYZ_2}, and is easily achievable given the large carrier Rabi frequency. We also note that the typical mean number of phonons after laser cooling in the resolved-sideband regime is $\bar{n}_n\sim 0.05$-$0.1$. 

Realizing a XYZ interaction according to our  scheme involves  two additional tones~\eqref{eq:3_tone}, requiring  a simple reprogramming of the AWG to provide the simultaneous frequencies $\delta\omega_\ell \in \{\omega_{\rm r},\omega_{\rm b},\omega_0,\omega_0+\Delta,\omega_0-\Delta\}$ along with the respective phases $\{\phi_{\rm r},\phi_{\rm b},0,0,0\}$. Alternatively, the AWG can be programmed to modulate the amplitude of the $\omega_0$ tone as $h_0(1+\xi \cos(\Delta t))$; such modulations have already been a key technique in  trapped-ion many-body spectroscopy~\cite{ions_spectroscopy}. While possible experimental limitations could include the sampling rate of the AWG ($>$ 1 GHz) and the response rate of the AOM ($>$20 MHz), these are both sufficiently fast to allow for the desired modulations ($\Delta/2\pi \approx 20$ kHz).

\section{ Numerical validation of the scheme}
\label{sec:num}
sec:num
 We start by visualising   $\langle\sigma_i^y(t)\rangle$ for a  two-ion setup with the parameters in Fig.~\ref{fig_num_validation}. The spin-phonon system is initialised in $\rho(0)=\ket{\psi_{0}}\!\!\bra{\psi_{0}}\otimes\rho_{\rm th}$, where $\ket{\psi_{0}}=\ket{+_y}\ket{-_y}$ is  the  spin state,    $\ket{\pm_y}$  are the  eigenstates of $\sigma^y$, and $\rho_{\rm th}$ is the vibrational thermal state  after laser cooling. In Fig.~\ref{fig_num_validation}{\bf (a)}, we compare the prediction of  the full  Hamiltonian $H(t)$ with the XYZ~\eqref{eq:XYZ} and the $h_0=0$ Ising model~\eqref{eq:Ising}, which clearly shows that the magnetisation exchange  is no longer described by the Ising model, but  instead by the  XYZ model. To quantify the accuracy, we  determine  how close the exact and effective  evolutions are, regardless of the possible initial states and   observables,  via the quantum channel fidelity 
$
\frak{F}(t):=\!\!\int\!\!{\rm d}\psi_{\rm s}\! \bra{\psi_{\rm s}}U_{\rm eff}^{\dagger}(t)\mathcal{E}\big(\!\ket{\psi_{\rm s}}\!\!\bra{\psi_{\rm s}}\!\big) U_{\rm eff}(t)\ket{\psi_{\rm s}},
$
 where  $\mathcal{E}\big(\!\ket{\psi_{\rm s}}\!\!\bra{\psi_{\rm s}}\!\big) :={\rm tr}_{\rm ph}\{U(t)\ket{\psi_{\rm s}}\!\!\bra{\psi_{\rm s}}\otimes\rho_{\rm th}U^\dagger(t) \}$, and one  integrates over the Haar measure  ${\rm d}\psi_{\rm s}$. This quantity can be  evaluated   efficiently~\cite{channel_fidelity}, although the numerics become considerably more demanding. Hence, we focus  on the XXZ model~\eqref{eq:coupling_constants} for $\xi=0$~\cite{comment_XXZ}, and  believe that the  results should be similar  for the XYZ case.   In Fig.~\ref{fig_num_validation}, we represent the  time-averaged error $\bar{\epsilon}=\frac{1}{t_{\rm f}}\int_0^{t_{\rm f}}{\rm d}t(1-\frak{F}(t))$, where $t_{\rm f}=\pi/J_{12}$,  as a function of: {\bf (b)} the spin phase $\phi_{\rm s}$ that controls the XXZ parameters~\eqref{eq:coupling_constants}, and {\bf (c)} the mean phonon number. These  results show that the accuracy of the XXZ  model in  representing the full spin-phonon dynamics is above 99$\%$ if $h_0/2\pi<$2kHz, for all phases $\phi_{\rm s}$, and for warm    phonons up to $\bar{n}_n<0.5$. In particular,  they  show that  phonon-induced errors  are negligible in the parameter regime~\eqref{eq:regime_XYZ_1}, which will allow us to study adiabatic protocols to  prepare  the XXZ model groundstate  by looking directly at the periodically-modulated spin model~\eqref{eq:time_dep_ising}. 
 
The  nearest-neighbour limit of the XXZ model hosts two different phases: a gapped antiferromagnetic   phase for $\phi_{\rm s}<\phi_{\rm s}^{\rm c}:=\cos^{-1}(1/\sqrt{3})$, and a gapless Luttinger liquid   for $\phi_{\rm s}>\phi_{\rm s}^{\rm c}$~\cite{xxz_luttinger}. The addition of frustration via next-to-nearest-neighbour interactions leads to a richer phase diagram with additional spontaneously dimerised~\cite{j1j2_model} and  gapless chiral~\cite{j1_j2_anisotropic} phases. The fate of these phases 
in presence of     long-range  interactions~\eqref{eq:XYZ} is an open  question, which could be  addressed with our setup if {\it (i)} preparation and {\it (ii)} detection of the groundstate are shown to be possible. We start   by discussing initialisation via adiabatic evolution and the role of the Hamiltonian symmetries for finite chains. For $N/2$ even, we choose $\ket{\psi_0}=\ket{-_y-_y-_y\cdots}$ , which can be prepared using global one-qubit gates, and approximates  the paramagnetic  groundstate of the Ising model~\eqref{eq:Ising} for $\phi_{\rm s}=0$, $\phi_{\rm d}=\pi/2$, and  $h_0\gg J_{ij}$. For $N/2$ odd, we choose $\ket{\psi_0}=\ket{-_y+_y-_y\cdots}$, 
which approximates the groundstate of a staggered  Ising model~\cite{comment_staggered}. One then   ramps down $h_0(t)\to 0$  slowly, such that the state  adiabatically follows  the groundstate of the Hamiltonian and ends in one of the Ising groundstates $\ket{\psi_p}=\frac{1}{\sqrt{2}}\left(1+p\otimes_{i=1}^N\sigma_i^z\right)\ket{+_x-_x+_x\cdots}$, where the parity $p=+1$ for $N/2$ even, and $p=-1$ for $N/2$ odd, is related to the $\mathbb{Z}_2$  symmetry of the Hamiltonian. This even-odd distinction is crucial for the rest of the protocol, where  the additional tones~\eqref{eq:3_tone} in Table~\ref{tab_1} are switched on, making sure that the constraints~\eqref{eq:regime_XYZ_1}-\eqref{eq:regime_XYZ_2} are fulfilled, and the spin phase  is slowly ramped up  to the desired value $\phi_{\rm s}(t)\to\phi_{\rm f}$. We  expect to have prepared the groundstate of the long-range  XXZ model~\eqref{eq:XYZ} corresponding to that particular $\phi_{\rm f}$, which has  parity $p=\pm 1$ for $N/2$ even/odd
as a consequence of the open boundary conditions of the finite chain~\cite{comment_odd}. In Fig.~\ref{fig_num_validation}{\bf (d)}, we represent the fidelity $\mathcal{F}_{\rm ad}(t_{\rm f},\phi_{\rm f})=|\braket{\epsilon_{\rm gs}^{\rm XXZ}(\phi_{\rm f})|\psi(t_{\rm f})}|^2$, where $\ket{\psi(t_{\rm f})}=\mathcal{T}\{\ee^{-\ii\int_0^{t_{\rm f} }{\rm d}\tau H(\tau)}\}\ket{\psi_0}$ is the  state evolving under the succession of the Ising~\eqref{eq:Ising} and  the periodically-modulated~\eqref{eq:time_dep_ising} spin Hamiltonians, according to the   previous sequence of adiabatic ramps, and $\ket{\epsilon_{\rm gs}^{\rm XXZ}(\phi_{\rm f})}$ is the exact groundstate of~\eqref{eq:XYZ}. We observe that the fidelity is very close to unity when the ramps are slow enough. For fast ramps, the fidelities are  compromised in the region $\phi_{\rm f}>\phi_{\rm s}^{\rm c}\approx 0.3\pi$, which is a consequence of a decrease in the energy gap, and would become accentuated as $N$ grows~\cite{kibble_zurek}. 

Once the desired ground states are adiabatically prepared, we can address the issue of detection. Trapped-ion experiments allow in-situ measurements of  spin-spin correlations  through  fluorescence~\cite{Ising_exp_ions}, or spectroscopic probes of  low-lying excitations~\cite{ions_spectroscopy}. The former would allow to distinguish between   Ising, Luttinger, dimerised, and spin-chiral orders discussed above, whereas the later would probe their gap.

\section{ Long-range  Heisenberg model} 
\label{sec:NLSM}
To get a flavour of the effect of  long-range interactions in our quantum simulator~\eqref{eq:XYZ}, we focus on the archetypical SU(2)-symmetric point   $\phi_{\rm s}=\phi_{\rm s}^{\rm c}$, $\xi=0$. In analogy to Haldane's result for the nearest-neighbour model~\cite{nlsm_heisenberg}, we map the low-energy properties of our  long-range Heisenberg Hamiltonian  onto  a non-linear sigma model (NLSM) described by the Lagrangian 
\beq
\mathcal{L}_{\rm NLSM}=\frac{1}{2g}(\partial_\mu\boldsymbol{\phi})\cdot(\partial^\mu\boldsymbol{\phi})
+\frac{\Theta}{8\pi}\epsilon^{\mu\nu}\boldsymbol{\phi}\cdot(\partial_\mu\boldsymbol{\phi}\times\partial_\nu\boldsymbol{\phi}).
\eeq
Here, $\boldsymbol{\phi}(x^\mu)$ is a three-component vector field associated to the staggered magnetisation,  constrained to  $|\boldsymbol{\phi}|^2=1$, and  defined on a  1+1 space-time $x^\mu=(vt,x)$. We have introduced the velocity $v=a_0{(\sum_{r} \tilde{J}_{2r-1} \chi)^{1/2}},$ where $a_0$ is the lattice spacing for bulk ions,  $\tilde{J}_{r}=4J_{i,i+r}$,  and  $\chi=\sum_{r} (-1)^{r+1} r^2\tilde{J}_{r}$. Additionally, we get a coupling constant $g=2(\sum_{r} \tilde{J}_{2r-1}\chi^{-1})^{1/2}/S,$  and a topological angle $\Theta=\pi$ (see Appendix~\ref{app:B}).

The  topological $\Theta$-term has drastic consequences on the NLSM. Under a renormalisation-group transformation, it   either flows  to the gapless fixed point of the  SU(2)$_1$  Wess-Zumino-Witten (WZW) conformal field theory when $g<g_{\rm c}$, or to a gapped fixed point for $g>g_{\rm c}$~\cite{nlsm_heisenberg_hamiltonian}. Truncating the long-range to  two neighbours, 
one finds $g=4/\sqrt{1-4\tilde{J}_2/\tilde{J}_1}$, such  that $g\to\infty$ as $\tilde{J}_2\to \tilde{J}_1/4$. This coincides roughly with the critical point towards the spontaneously dimerised phase of the $J_1$-$J_2$ Heisenberg model~\cite{j1j2_model}, which is  a truncation  of the lattice version  of the  SU(2)$_1$  WZW theory: the Haldane-Shastry model~\cite{Haldane_shastry,WZW_haldane_shastry}. Therefore, the  NLSM mapping identifies the WZW critical point with the instability  $g_{\rm c}\to\infty$.

\begin{figure}
\centering
\includegraphics[width=0.9\columnwidth]{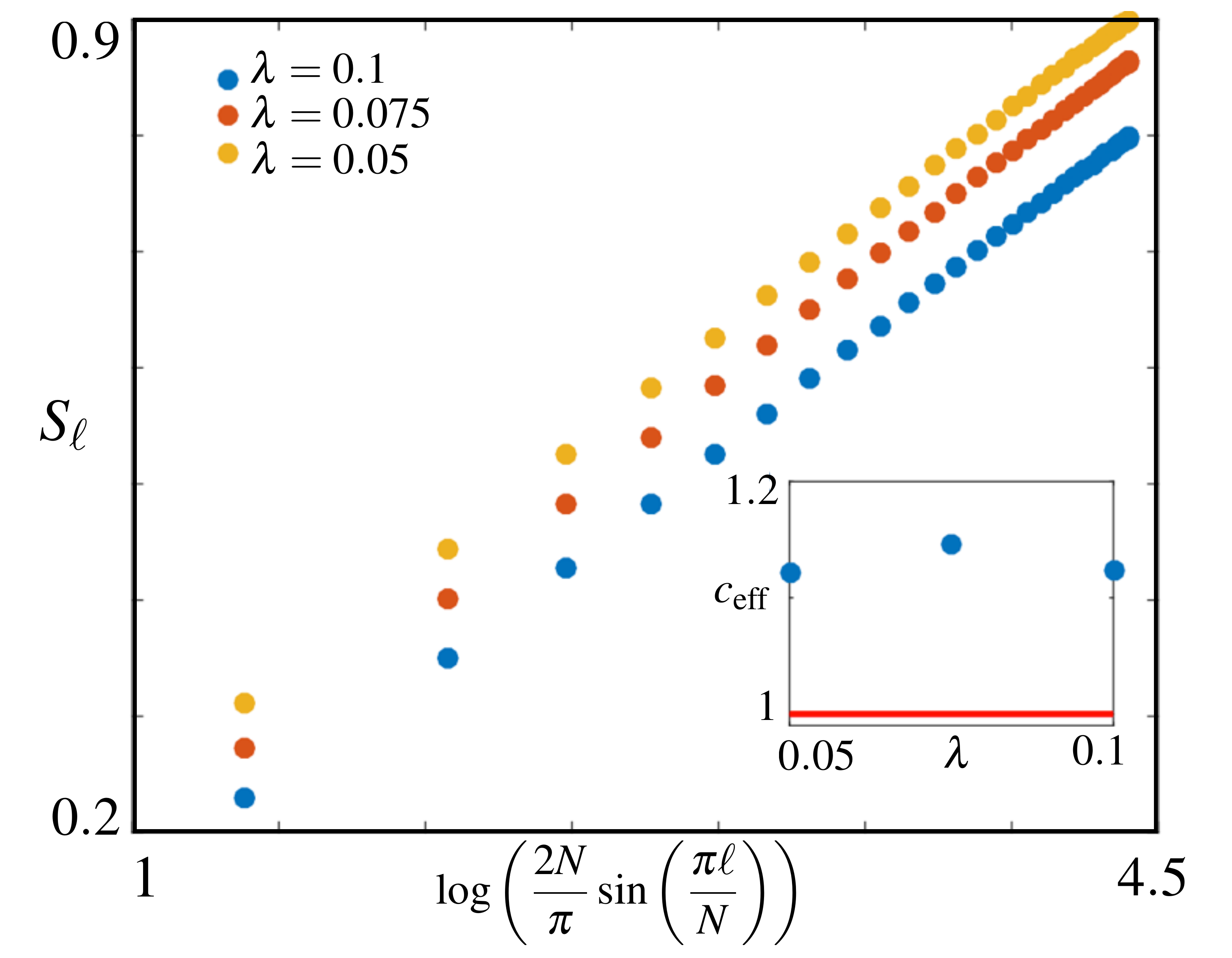}
\caption{ {\bf Central charge in the long-range Heisenberg model:} Entanglement entropies $S_\ell$ of blocks of length $\ell$ in a chain of $N=128$ spins computed with a Matrix-Product-State ansatz for the ground state of the long range Heisenberg model with couplings~\eqref{eq:couplings_main} using bond dimensions up to 200 to guarantee the convergence of the ground state energy up to 12 digits. The ground state is obtained using the time-dependant variational algorithm originally presented in \cite{juto1}  and later reformulated for finite chains with long range interactions in \cite{luca1, luca2}. The entanglement entropies, for large enough $\ell$  scale linearly with the variable $y_\ell=\log(\frac{2 N}{\pi} \sin(\frac{\pi \ell}{N}))$,  as expected for a conformally-invariant spin chain, for all the values considered $\lambda\in\{0.1, 0.075, 0.05\}$. Quantitatively, the pre-factor of the scaling  exceeds the  central charge $c=1$ of the nearest-neighbor model (red line). The  coefficients $c_{\rm eff}$ are obtained by a linear fit of the larger $\ell$ entropies to $S_\ell=\frac{c_{\rm eff}}{6}y_\ell+a$, and shown in the inset. }
\label{fig_ent_entropy}
\end{figure}

For the full long-range model~\eqref{eq:XYZ}, the distance-decay of the spin-spin interactions in the trapped-ion crystal is
\beq
\label{eq:couplings_main}
\tilde{J}_r\approx 4\frac{|J_0\lambda|}{|r|^3}+8|J_0|\big({\rm sgn}(\lambda)\big)^{1+r}\ee^{-\frac{|r|a_0}{\xi_0}}\theta(r-2),
\eeq
where $J_0$  quantifies the spin couplings, $\lambda$ determines how close the MS forces lie from the vibrational sidebands, and $\xi_0\approx-a_0/\log(|\lambda|)$ for far-detuned forces $|\lambda|\ll 1$(see Appendix~\ref{app:C}). We  find that $g\approx\sqrt{7\zeta(3)/2+2\lambda}/\sqrt{\log 2-2\lambda}$, where $\zeta(n)$ is the Riemann zeta function. For $|\lambda|\ll 1$, the NLSM coupling is thus finite, and no instability takes place. Therefore, the NLSM predicts that our long-range Heisenberg model  flows to the  WZW  fixed point (i.e. gapless, power-law  correlations, and logarithmic scaling of the entanglement entropy), instead of the  spontaneously dimerised gapped phase. 

We test   this prediction numerically by  a Matrix-Product-state  calculation of the ground-state entanglement entropy $S_\ell=-{\rm Tr}\{\rho_\ell\log\rho_\ell\}$, where $\rho_\ell={\rm Tr}_{N-\ell}\{\ket{\epsilon_{\rm g}}\bra{\epsilon_{\rm g}}\}$ is the reduced density matrix for a block of $\ell$ sites inside a chain of length $N$. In Fig.~\ref{fig_ent_entropy}, $S_\ell$ is depicted for different block sizes as 
 a function of $y_\ell=\log(\frac{2 N}{\pi} \sin(\frac{\pi \ell}{N}))$, and considering  different  interaction ranges~\eqref{eq:couplings_main}. For a chain with open boundary conditions, conformal field theory predicts   $S_\ell=\frac{c}{6}y_\ell + a$, where $c$ is the central charge, and $a$ is a non-universal constant~\cite{cft_entanglement_entropy}. Our numerical results display the predicted linear scaling of a gapless phase but, interestingly,  long range interactions change quantitatively the value of the central charge, which gets promoted from the short-range prediction $c=1$ to  an effective larger value $c_{\rm eff}>1$~\cite{comment_c_nn} (see inset in Fig.~\ref{fig_ent_entropy}). A similar behavior  had already been observed in the long-range quantum Ising model at criticality~\cite{luca1}, and we provide a possible qualitative explanation of both below.

This effect can be understood qualitatively from the scaling of $S_{\ell}$ in closed spin chains with a defect~\cite{cft_entanglement_entropy,affleck_impurity}, namely a weaker spin-spin coupling $0\le J_{\rm d}\le J$ in a particular  bond.  Depending on the particular model, one can find $S_\ell=\frac{c_{\rm eff}}{6}y_{\ell} +a'$ with $c_{\rm eff}$ varying continuously with $J_{\rm d}$ between the open- and closed-chain limits $c \le c_{\rm eff} \le 2c$. Our numerics show a similar behavior, which  can be understood intuitively by noticing that  long-range interactions induce direct couplings between the otherwise non-interacting boundaries of the chain, 
yielding  a hybrid between  the closed- and open-chain limits. Accordingly, one would expect $ c_{\rm eff}>c$, which is the result found in the inset of Fig.~\ref{fig_ent_entropy}. 

\section{ Conclusions and outlook}
\label{sec:conc}
 We have proposed a realistic  QS of long-range Heisenberg-type models based on quasi-periodically driven trapped ions. Making use of a single branch of phonons, this scheme is readily applicable to existing experiments that simulate the  Ising model in either Paul~\cite{Ising_exp_ions} or Penning~\cite{Ising_exp_penning} traps. Since the Heisenberg model describes the magnetic properties of  Mott-insulating materials, our work opens the possibility of  using trapped-ion quantum simulators to assess real-world problems.   We have presented analytic and numerical evidence that the 1D long-range model shares similar topological properties  with the paradigmatic nearest-neighbour limit~\cite{nlsm_heisenberg}. It would be very interesting to generalise these methods to ladders with triangular motifs, as these arise naturally in the experiment, and  may provide an alternative to observe non-trivial effects~\cite{nlsm_ladders} that appear in integer-spin Heisenberg models~\cite{nlsm_heisenberg,s_1_heisenberg_trapped_ions,s_1_long_range_Heisneberg}.

\acknowledgements
A.B. and G.S. acknowledge support from Spanish MINECO Projects FIS2015-70856-P, FIS2015-69167-C2-1-P, and CAM regional research consortium QUITEMAD+.  P.R. acknowledges support from the U.S. AFOSR award no. FA9550-16-1-0277.

\appendix

\section{Derivation of the effective long-range XYZ model}
\label{app:A}
In this Appendix, we present a detailed derivation of the effective Heisenberg-type XYZ model~\eqref{eq:XYZ} directly from the Hamiltonian of the periodically-driven trapped-ion chain $H(t)=H_0+H_{\rm MS}+H_{\rm C,3}$ in Eqs.~\eqref{eq:MS_force}, and~\eqref{eq:3_tone}.
For the shake of completeness, however, we note that: {\it (i)} the un-driven trapped-ion crystal is described by $H_0$ in the harmonic approximation, which is valid for low-enough temperatures and small vibrations around the equilibrium positions~\cite{normal_modes}. {\it (ii)} The M$\o$lmer-S$\o$rensen force $H_{\rm MS}$~\eqref{eq:MS_force} is obtained from the light-matter interaction of a pair of laser beams~\cite{molmer_sorensen} coupled to the internal transition in the regime of resolved sidebands~\cite{wineland_review}, such that the frequency of one laser is tuned to the first red sideband $\omega_{\rm r}=\omega_0-\omega_{n,x}+\delta_n,$ whereas the other laser excites the first blue sideband $\omega_{\rm b}=\omega_0+\omega_{n,x}-\delta_n$. These excitations are in practice  generated by passing a laser beam through an acousto-optic modulator driven by an arbitrary waveform generator (see Fig.~\ref{fig_scheme_sm_1}{\bf (a)}). When the opposite detunings fulfil $\delta_n\ll\omega_{n,x}$, the light matter interaction yields Eq.~\eqref{eq:MS_force}, where 
the light forces and zero-point displacement can be expressed as
\beq
\label{eq:forces}
\mathcal{F}_{in}=\ii \frac{\Omega_{\rm L}\Delta {\bf k}\cdot {\bf e}_x}{2}\sqrt{\frac{\omega_x}{\omega_{n,x}}}\mathcal{M}_{i,n}\ee^{-\half\sum_n\mathcal{M}_{in}^2\eta_n^2},\hspace{3ex} x_0=\frac{1}{\sqrt{2m \omega_x}}
\eeq
in terms of the common Rabi frequency (Lamb-Dicke parameter) $\Omega_{\rm L}\ll \omega_{n,x}\ll \omega_0$ ($\eta_n=\Delta {\bf k}\cdot {\bf e}_x/\sqrt{2m\omega_{n,x}}\ll 1$) of both MS laser beams, the trap $\omega_x$ and normal-mode $\omega_{n,x}$  frequencies, and the corresponding normal-mode displacements $\mathcal{M}_{i,n}$. In Eq.~\eqref{eq:MS_force}, we have  written explicitly the phases of the red- and blue-sideband beams $\phi_{\rm r},\phi_{\rm b}$  (see Fig.~\ref{fig_scheme_sm_1}{\bf (a)}). {\it (iii)} The carrier driving  $H_{\rm C,3}$~\eqref{eq:3_tone} can be obtained  from the light-matter interaction with an additional laser beam that pases through the acousto-optic modulator producing three tones with frequencies
 $\omega_{\rm d, 1}=\omega_0$, $\omega_{\rm d, 2}=\omega_0+\Delta$, $\omega_{\rm d, 3}=\omega_0-\Delta$, such that $\Delta\ll\omega_{n,x}$, which can be also included in the waveform generator. In this later case, the strengths of the carrier drivings are limited to $h_{\rm t}=\Omega_{\rm d, t}/2\ll\omega_{n,x}/\eta_n$~\cite{wineland_review}. By virtue of the arbitrary wave generator, it is possible to  control not only the driving strengths, but also the phases of the drivings $\phi_{\rm d, t}$ with respect to $\phi_{\rm r}$, and $\phi_{\rm b}$, which allows us to consider the values listed in Table~\ref{tab_1}.

Let us now start with the derivation of the effective XYZ  model. The first step is to move to the interaction picture with respect to $H_0$, $\ket{\tilde{\psi}}=\ee^{\ii H_0 t}\ket{\psi}$, such that $\ii\partial_t \ket{\tilde{\psi}}=(\tilde{H}_{\rm MS}(t)+\tilde{H}_{\rm C,3}(t))\ket{\tilde{\psi}}$, with the following drivings
\beq
\tilde{H}_{\rm MS}(t)=\sum_{i,n}\mathcal{F}_{in}x_0\sigma_i^{\phi_{\rm s}}(a_{n,x}^{\phantom{\dagger}}\ee^{\ii\phi_{\rm m}-\ii \delta_n t}+a^\dagger_{n,x}\ee^{-\ii\phi_{\rm m}+\ii \delta_n t}),
\eeq
where in addition to the average spin phase $\phi_{\rm s}$, and the spin operator $\sigma_i^{\phi_{\rm s}}$ introduced below Eq.~\eqref{eq:Ising} in the main text, also the relative motional phase  $\phi_{\rm m}:=(\phi_{\rm r}-\phi_{\rm b})/2$ appears. In this interaction picture, the carrier driving becomes
\beq
\label{eq:carr_driving}
\tilde{H}_{\rm C,3}(t)=\sum_i h_0\big(1+\xi\cos(\Delta t)\big)\sigma_i^x,
\eeq
where we have used the parameters listed in Table~\ref{tab_1}, such that $\xi$ depends on the ratio of the Rabi frequencies of the $\pm \Delta$ detuned tones with respect to the resonant one. Let us now perform the following unitary transformation $ \ket{\hat{\psi}}=\hat{U}(t)\ket{\tilde{\psi}}$, with
\beq
\hat{U}(t)=\ee^{\ii\sum_i h_0\big(t+\frac{\xi}{\Delta}\sin(\Delta t)\big)\sigma_i^x},
\eeq
such that the transformed state evolves only under the transformed MS force $\ii\partial_t \ket{\hat{\psi}}=\hat{H}_{\rm MS}(t)\ket{\hat{\psi}}$, namely
\beq
\label{eq:MS_driven_force}
\hat{H}_{\rm MS}(t)=\sum_{i,n}\mathcal{F}_{in}x_0\sigma_i^{\phi_{\rm s}}(t)(a_{n,x}^{\phantom{\dagger}}\ee^{\ii\phi_{\rm m}-\ii \delta_n t}+a^\dagger_{n,x}\ee^{-\ii\phi_{\rm m}+\ii \delta_n t}).
\eeq
Here, we have introduced the transformed spin operator $\sigma_i^{\phi_{\rm s}}(t):=\hat{U}(t)\sigma_i^{\phi_{\rm s}}\hat{U}(t)^\dagger=\cos\phi_{\rm s}\sigma_i^x-\sin\phi_{\rm s}\hat{U}(t)\sigma_i^y\hat{U}(t)^\dagger$, which can be expressed in terms of the $m$-th order Bessel functions $\frak{J}_m(x)$ of the  first class as follows 
\begin{equation}
\begin{split}
\label{eq:mod_pauli}
\sigma_i^{\phi_{\rm s}}(t)=\cos\phi_{\rm s}\sigma_i^x -&\sin\phi_{\rm s}\sum_{m\in\mathbb{Z}}\frak{J}_m\left(\xi\frac{h_0}{\Delta}\right)\cos\big((2h_0+m\Delta)t\big)\sigma_i^y\\
+&\sin\phi_{\rm s}\sum_{m\in\mathbb{Z}}\frak{J}_m\left(\xi\frac{h_0}{\Delta}\right)\sin\big((2h_0+m\Delta)t\big)\sigma_i^z.
\end{split}
\end{equation}

Let us remark that the effect of the carrier  driving~\eqref{eq:carr_driving} on  the full light-matter interaction of the MS scheme  is to introduce a comb of new frequencies $\nu_{\pm, m}:=\pm(2 h_0+m\Delta)$, where $m\in\mathbb{Z}$ (see Fig.~\ref{fig_scheme_sm_1}{\bf (b)}). Hence, the validity of the description of such a light-matter   coupling in terms of the periodically-modulated MS force~\eqref{eq:MS_driven_force} rests on the condition that none of these frequencies hits a  resonance with the carrier or any higher-order sideband in the original light-matter interaction, or that in the event of such a resonance, the coupling strength gets suppressed in comparison with the aforementioned MS force. This imposes the following constraints on the carrier driving parameters
\beq
\label{eq:higher_sidebands}
h_0\sim\Delta\ll\omega_{n,x}\sim\omega_x, \hspace{2ex}\xi<\half.
\eeq
The first inequality guarantees that the resonance will only occur for a very large number of `photons'  $m\gg 1$ absorbed from the drive (see Fig.~\ref{fig_scheme_sm_1}{\bf (b)}), whereas the second inequality guarantees that the strength of such a resonance gets exponentially suppressed with respect to the strength of the MS force as $m$ increases  (see Fig.~\ref{fig_scheme_sm_1}{\bf (c)}).

\begin{figure*}
\centering
\includegraphics[width=1.9\columnwidth]{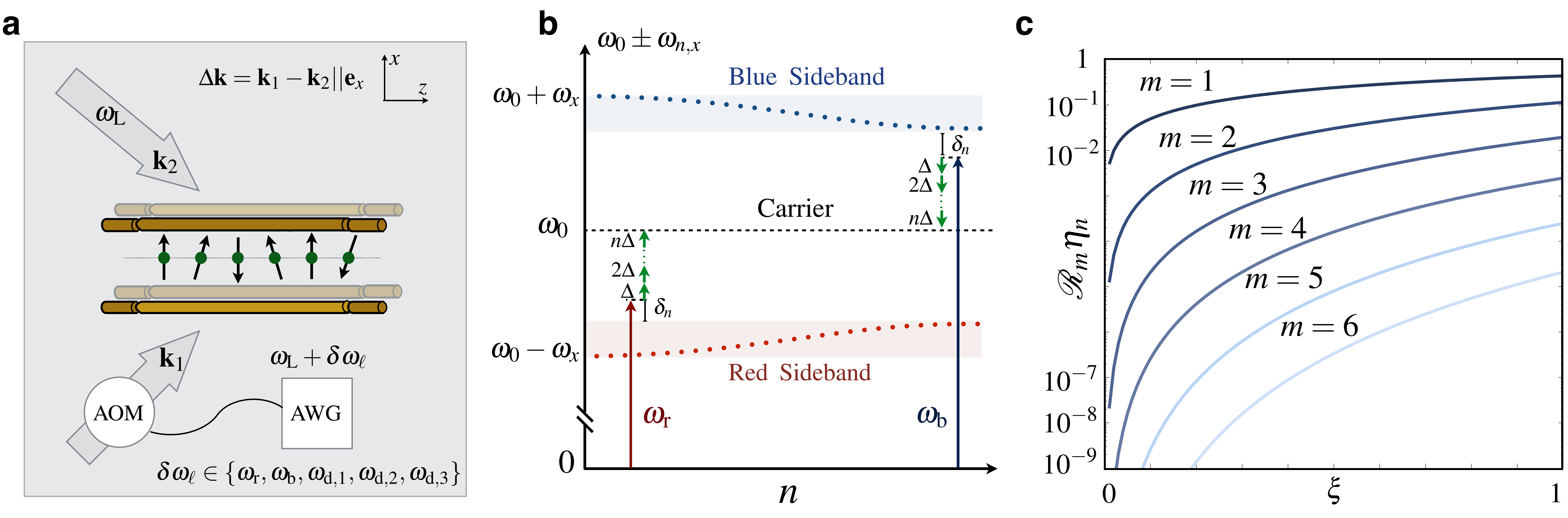}
\caption{ {\bf Scheme of the quasi-periodically modulated Molmer-Sorensen force:} {\bf (a)} Ions (green circles) forming a chain within the electrodes of a linear Paul trap. Two internal electronic levels of each ion form a pseudospin (thin arrows) that is coupled to the incident laser beams (wide arrows), the beatnote of which is controlled by an acusto-optical modulator (AOM) through an arbitrary wave generator (AWG). By using the different modulation frequencies in the AWG listed in Table~\ref{tab_1}, we obtain a quasi-periodically modulated Molmer-Sorensen force that couples the spins to the motion along the axis ${\bf e}_x||\Delta {\bf k}$. {\bf (b)} Scheme for the blue and red sidebands ($\omega_0\pm\omega_{n,x}$) of the whole vibrational branch for the Molmer-Sorensen beams, and their symmetric detuning $\delta_n$. We also represent with small arrows the comb of frequencies $m\Delta$, with $m\in\mathbb{Z}$, due to the additional drivings introduced in the scheme~\eqref{eq:mod_pauli}, which could hit an undesired resonance (e.g. carrier transition). {\bf (c)} Ratio of the coupling strength $\mathcal{R}_m$ between  the possible spurious resonance with the carrier due to the additional comb of frequencies $m\Delta$, and  the desired MS sideband,  as a function of  $m\in\{1,\cdots, 6\}$. We see that for the high resonances (i.e. high $m$) required by the constraint~\eqref{eq:higher_sidebands}, lead to a vanishingly small ratio, such that these terms can be safely neglected for the timescales of interest of the experiment.}
\label{fig_scheme_sm_1}
\end{figure*}

Provided that these conditions are met, we now start from Eq.~\eqref{eq:MS_driven_force} to prove that the time-evolution operator in this transformed basis  $\hat{U}(t)=\mathcal{T}\{\ee^{-\ii\int_0^t{\rm d}\tau \hat{H}_{\rm MS}(\tau)}\}$ gives raise to the desired effective XYZ Hamiltonian $\hat{U}(t)\approx \ee^{-\ii H_{\rm XYZ} t}$ in Eq.~\eqref{eq:XYZ}. We use the so-called Magnus expansion~\cite{magnus} to second order
\begin{widetext}
\beq
\label{eq:magnus_exp}
\hat{U}(t)\approx\ee^{\Omega_1(t)+\Omega_2(t)},\hspace{2ex} \Omega_1(t)=-\ii\int_0^t {\rm d}\tau \hat{H}_{\rm MS}(\tau), \hspace{2ex}\Omega_2(t)=-\frac{1}{2}\int_0^t {\rm d}\tau_1\int_0^{\tau_1} {\rm d}\tau_2 [ \hat{H}_{\rm MS}(\tau_1), \hat{H}_{\rm MS}(\tau_2)].
\eeq
\end{widetext}
Integrating by parts, we find the following expression for the first-order contribution
\begin{widetext}
\beq
\Omega_1(t)=\sum_{i,n}\frac{\mathcal{F}_{in} x_0}{\delta_n}a_n\left(\sigma_i^{\phi_{\rm s}}(t)\ee^{\ii(\phi_{\rm m}-\delta_n t)}-\sigma_i^{\phi_{\rm s}}(0)\ee^{\ii\phi_{\rm m}}	\right)-\int_0^t{\rm d}\tau\frac{\mathcal{F}_{in} x_0}{\delta_n}\frac{{\rm d}\sigma_i^{\phi_{\rm s}}}{{\rm d}\tau} a_n\ee^{\ii(\phi_{\rm m}-\delta_n \tau)}-{\rm H.c.},
\eeq
\end{widetext}
where the integral in the right hand side involves the derivative of the driven Pauli operator~\eqref{eq:mod_pauli}. This first-order contribution to the unitary evolution can be understood as a spin-dependent displacement acting on the phonons, and leading to spin-phonon correlations that compromise the validity of an effective spin model. As such, these terms must be minimised, which requires  avoiding the possible new resonances brought up by the frequency comb of the driven Pauli operator~\eqref{eq:mod_pauli}. This can be guaranteed by imposing a more restrictive constraint than Eq.~\eqref{eq:higher_sidebands} on the carrier parameters 
\beq
\label{eq:const_driving}
h_0\sim\Delta\ll\delta_{n}\ll\omega_x, \hspace{2ex}\xi<\half.
\eeq
Then, the integral in the right hand side is suppressed with respect to the left hand side, and 
\beq
\Omega_1(t)\approx\sum_{i,n}\frac{\mathcal{F}_{in} x_0}{\delta_n}a_n\left(\sigma_i^{\phi_{\rm s}}(t)\ee^{\ii(\phi_{\rm m}-\delta_n t)}-\sigma_i^{\phi_{\rm s}}(0)\ee^{\ii\phi_{\rm m}}	\right)-{\rm H.c.},
\eeq
where we have neglected terms of the order $\mathcal{O}\left(\frac{\mathcal{F}_{in} x_0}{\delta_n}\frac{h}{\delta_n}\right)$.
In order to make  $\Omega_1(t)\approx 0$, we must thus work in the far-detuned regime of the MS force
\beq
\label{eq:const_force}
\mathcal{F}_{in} x_0\sqrt{1+\bar{n}_n}\ll \delta_n,
\eeq
where $\bar{n}_n$ stands for the mean number of phonons of the ion chain after laser cooling has been performed. Hence, provided that the constraints in Eqs.~\eqref{eq:const_driving} and~\eqref{eq:const_force} are fulfilled, the dynamics of the spins will be governed by the second-order contribution in Eq.~\eqref{eq:magnus_exp}. This contribution can introduce terms that do not oscillate periodically in time and  would thus not be negligible under the constraint~\eqref{eq:const_force},  but instead lead to  an effective Hamiltonian. Once again, the integration by parts is of practical importance to identify the leading-order terms in the regime of Eq.~\eqref{eq:const_driving}, and we find 
\beq
\Omega_2(t)\approx-\ii\!\int_0^t\!\!{\rm d}\tau\!\left(\!\sum_{i>j}\!J_{ij}\sigma_i^{\phi_{\rm s}}(\tau)\sigma_j^{\phi_{\rm s}}(\tau)+\sum_{i,n}\lambda_{in}\sigma_i^x(2a_{n,x}^{\dagger}a_{n,x}^{\phantom{\dagger}}+1)\!\!\right),
\eeq
where we have neglected $\mathcal{O}\left(J_{ij}\frac{h}{\delta_n}\right)$ terms, and introduced the spin-spin couplings $J_{ij}$ defined below Eq.~\eqref{eq:Ising} of the main text, 
\beq
\label{eq:spin-spin_int}
J_{ij}=-\sum_n\frac{\mathcal{F}_{in}^{\phantom{*}}\mathcal{F}_{jn}^*x_0^2}{\delta_n}+{\rm c.c.},
\eeq
and some residual spin-phonon coupling strength
\beq
\lambda_{in}=\sin^2\phi_{\rm s}(\mathcal{F}_{in}x_0)^2\sum_{m\in\mathbb{Z}}\frak{J}_m^2\left(\frac{h_0}{\Delta}\xi\right)\frac{m\Delta+2h_0}{\delta_n^2-(m\Delta+2h)^2}.
\eeq
Since we have constrained the driving parameters to the regime in Eq.~\eqref{eq:const_driving}, and we have in particular that  $h_0\sim\Delta$ and $\xi<\half$, the $m$-photon resonances on the previous equation are exponentially suppressed as $m$ increases, and we
can approximate  $\lambda_{in}=\sin^2\phi_{\rm s}(\mathcal{F}_{in}x_0)^2\frak{J}_0^2\left(\frac{h_0}{\Delta}\xi\right)2h_0/\delta_n^2\sim \mathcal{O}\left(J_{ii}(1+2\bar{n}_n)\sin^2\phi_{\rm s} \cdot h_0/\delta_n\right)$. According to this discussion, and provided that $h_0\ll \delta_n$~\eqref{eq:const_driving}, this residual spin-phonon coupling becomes negligible, and we obtain an effective spin model described by a  periodically-modulated Ising Hamiltonian
\beq
\tilde{U}(t)\approx\mathcal{T}\{\ee^{-\ii\int_0^t{\rm d}\tau H_{\rm eff}(\tau)}	\},\hspace{2ex} H_{\rm eff}(\tau)=\sum_{i>j}J_{ij}\sigma_i^{\phi_{\rm s}}(\tau)\sigma_j^{\phi_{\rm s}}(\tau).
\eeq
In the numerical simulations presented in the main text, we explore what particular values of $h_0,\Delta,$ and $\xi$, fulfilling the above constraints, lead to a negligible spin-phonon coupling while simultaneously allowing for a wide tunability of the  spin model.

The remaining task is to demonstrate that such a periodically-modulated Hamiltonian leads to the desired XYZ  model. So far, the derivation has only imposed that the strength and frequency of the carrier driving~\eqref{eq:carr_driving} must have the same order of magnitude $h_0\sim \Delta$. We will now show that by imposing a particular ratio $h_0/\Delta$, it is possible to engineer the aforementioned XYZ model. Instead of using Eq.~\eqref{eq:mod_pauli}, it will prove simpler to introduce the states $\ket{\pm_i}=(\ket{{\uparrow}_i}\pm\ket{\downarrow_i})/\sqrt{2}$, such that
\beq
\sigma_i^{\phi_{\rm s}}(t)=\cos\phi_{\rm s}\sigma_i^x-\sin\phi_{\rm s}\left(\ii\ee^{\ii 2h_0\left(t+\frac{\xi}{\Delta}\sin(\Delta t)\right)}\ket{+_i}\bra{-_i}+{\rm H.c.}\right).
\eeq
We now substitute in the periodically-modulated spin model, and obtain
$H_{\rm eff}(t)=\sum_{i>j} (h_{ij}^{(1)}+h_{ij}^{(2)}+h_{ij}^{(3)}),$ where
\begin{widetext}
\beq
\begin{split}
h_{ij}^{(1)}&:=J_{ij}\cos^2\phi_{\rm s}\sigma_i^x\sigma_j^x,\\
h_{ij}^{(2)}&:=J_{ij}\cos\phi_{\rm s}\sin\phi_{\rm s}\left(\sigma_i^x\left(\ii\ee^{\ii 2h_0\left(t+\frac{\xi}{\Delta}\sin(\Delta t)\right)}\ket{+_j}\bra{-_j}+{\rm H.c.}\right)+\left(\ii\ee^{\ii 2h_0\left(t+\frac{\xi}{\Delta}\sin(\Delta t)\right)}\ket{+_i}\bra{-_i}+{\rm H.c.}\right)\sigma_j^x\right),\\
h_{ij}^{(3)}&:=J_{ij}\sin^2\phi_{\rm s}\left(\ii\ee^{\ii 2h_0\left(t+\frac{\xi}{\Delta}\sin(\Delta t)\right)}\ket{+_i}\bra{-_i}+{\rm H.c.}\right)\left(\ii\ee^{\ii 2h_0\left(t+\frac{\xi}{\Delta}\sin(\Delta t)\right)}\ket{+_j}\bra{-_j}+{\rm H.c.}\right).
\end{split}
\eeq
\end{widetext}
In the contribution of $h_{ij}^{(2)}$, we obtain again the frequency comb in terms of multiples of the driving frequency, which contribute with terms that oscillate in time as follows $\sum_{m\in\mathbb{Z}}J_{ij}\frak{J}_m\big(2\xi\frac{h_0}{\Delta}\big)\ee^{\ii(2h_0+m\Delta)t}$. If we impose
\beq
\label{eq:constraint_spin_J}
{\rm max}\{J_{ij}\}\ll 2h_0=\frac{\Delta}{2}
\eeq
all these terms can be neglected using a rotating wave approximation, such that $h_{ij}^{(2)}\approx 0$. The situation for $h_{ij}^{(3)}$ is different, as this term can be rewritten  as follows
\begin{widetext}
\beq
h_{ij}^{(3)}=J_{ij}\sin^2\phi_{\rm s}\left(\ket{+_i}\bra{-_i}\cdot\ket{-_j}\bra{+_j}-\sum_{m\in\mathbb{Z}}\frak{J}_m\left(4\xi\frac{h_0}{\Delta}\right)\ee^{\ii(4h_0+m\Delta)t}\ket{+_i}\bra{-_i}\cdot\ket{+_j}\bra{-_j}+{\rm H.c.}\right).
\eeq
\end{widetext}
One thus finds that, under the constraints~\eqref{eq:constraint_spin_J}, several terms can be neglected under a rotating-wave approximation, except for  certain of resonances that may be considered as a spin analogue of   the photon-assisted tunnelling resonances  in periodically-modulated quantum systems~\cite{pat}. We thus obtain
\begin{widetext}
\beq
h_{ij}^{(3)}\approx J_{ij}\sin^2\phi_{\rm s}\bigg(\ket{+_i}\bra{-_i}\cdot\ket{-_j}\bra{+_j}-\frak{J}_{-1}(\xi)\ket{+_i}\bra{-_i}\cdot\ket{+_j}\bra{-_j}+{\rm H.c.}\bigg).
\eeq
\end{widetext}
Remarkably enough, we get an effective time-independent Hamiltonian that can be rewritten as follows
\beq
H_{\rm eff}(t)\approx H_{\rm XYZ}=\sum_{i>j} \left(J_{ij}^x\sigma_i^{x}\sigma_j^{x}+J_{ij}^y\sigma_i^{y}\sigma_j^{y}+J_{ij}^z\sigma_i^{z}\sigma_j^{z}\right),
\eeq
where the different coupling constants have been written in Eq.~\eqref{eq:coupling_constants} of the main text. If we consider the additional unitaries that were used to transform to the actual basis, we have shown that the full time-evolution of the driven trapped-ion crystal can be expressed as $U(t)\approx\ee^{-\ii t H_0}\ee^{-\ii\sum_i h_0\big(t+\frac{\xi}{\Delta}\sin(\Delta t)\big)\sigma_i^x}\ee^{-\ii t H_{\rm XYZ}}=:U_0(t)\ee^{-\ii t H_{\rm eff}}$, as used in the main text of this article.

\section{Non-linear sigma model for a long-range Heisenberg Hamiltonian}
\label{app:B}

In this Appendix, we present a detailed derivation of the mapping between the long-range anti-ferromagnetic Heisenberg model (LRHM), and  the O(3) non-linear sigma model (NLSM). The LRHM is a lattice model of interacting spins obtained from Eqs.~\eqref{eq:XYZ} and~\eqref{eq:coupling_constants} after setting $\phi_{\rm s}=\phi_{\rm s}^{\rm c}=\cos^{-1}(1/\sqrt{3})$ and $\xi=0$, and described by the SU(2)-invariant spin Hamiltonian
\beq
\label{eq:LRHM}
H_{\rm LRHM}=\sum_{i>j} \tilde{J}_{ij}\boldsymbol{S}_i\cdot\boldsymbol{S}_j,\hspace{3ex} \boldsymbol{S}_i=\frac{1}{2}( \sigma_i^{x}, \sigma_i^{y},\sigma_i^{z}),
\eeq
where we have introduced $\tilde{J}_{ij}=4J_{ij}>0$. The NLSM is  a relativistic quantum field theory in a 1+1 space-time $x^\mu=(vt,x)$ for a vector field $\boldsymbol{\phi}(x^\mu)$  on a 2-sphere, described by the  Lagrangian density
\beq
\label{eq:NLSM}
\mathcal{L}_{\rm NLSM}=\frac{1}{2g}(\partial_\mu\boldsymbol{\phi})\cdot(\partial^\mu\boldsymbol{\phi})
+\frac{\Theta}{8\pi}\epsilon^{\mu\nu}\boldsymbol{\phi}\cdot(\partial_\mu\boldsymbol{\phi}\times\partial_\nu\boldsymbol{\phi}), 
\eeq
where $|\boldsymbol{\phi}(x^\mu)|^2=1$. Here, $g>0$ is the coupling constant, $\epsilon^{\mu\nu}$ is the Levi-Civita symbol,  $\Theta$ is the so-called topological angle, and repeated indices are summed. By performing a Wick rotation $vt\to-\ii\tau$,  the action associated to this Lagrangian is finite if  $\lim_{|\boldsymbol{x}|\to\infty} \boldsymbol{\phi}(\boldsymbol{x})=\boldsymbol{\phi}_0$. Hence,  all values of the fields at infinity are the same $\boldsymbol{\phi}_0$, and the Euclidean space-time becomes isomorphic to a 2-sphere~\cite{fradkin_book}. In this case,  $\boldsymbol{\phi}(\boldsymbol{x})$ can be considered as a mapping of the 2-sphere of the compactified space-time onto the 2-sphere of the vector fields, and  the  Euclidean action can be written as $S=\int {\rm d}^2x \mathcal{L}_{\rm NLSM}=-\int {\rm d}^2x\left(\frac{1}{2g}(\partial_\mu\boldsymbol{\phi})^2+\ii\Theta W\right)$, where $W=\frac{1}{8\pi}\int {\rm d}^2x\epsilon^{\mu\nu}\boldsymbol{\phi}\cdot(\partial_\mu\boldsymbol{\phi}\times\partial_\nu\boldsymbol{\phi})\in\mathbb{Z}$ is the winding number of the mapping. Since the action will be exponentiated in a path-integral approach, and $W\in\mathbb{Z}$, one directly sees that $\Theta$ is defined modulo $2\pi$, and can be thus interpreted as an angle: a topological angle that controls the appearance of such a topological term in the action, and leads to important non-perturbative effects. For models invariant under parity $x^1\to-x^1$, the invariance of the action imposes that $\Theta\in\{0,\pi\}$, which is responsible for the massive/massless character of the low-energy excitations of the NLSM, respectively.

In the nearest-neighbor limit of the LRHM~\eqref{eq:LRHM}, Haldane showed that a mapping between both models exists and becomes exact in the  large-$S$ limit, where the spin-1/2 operators~\eqref{eq:LRHM} are substituted by spin-$S$ operators  that  also fulfil the $\mathfrak{su}(2)$ algebra $[S_i^a,S_j^b]=\ii\epsilon^{abc}S_i^c\delta_{i,j}$, but have magnitude $\boldsymbol{S}_i^2=S(S+1)$~\cite{nlsm_heisenberg}. The mapping leads to $g=2/S$, $v=2\tilde{J}Sa$, and $\Theta=2\pi S\hspace{0.2ex}{\rm mod}(2\pi)$, where  $a$ is the  lattice spacing. Accordingly, the topological angle vanishes $\Theta=0$ for {\it integer spin}, and  one recovers the standard  NLSM without a topological term, which  displays massive excitations, as shown by perturbative renormalization-group arguments for weak couplings $g\ll 1$~\cite{polyakov_rg_On,fradkin_book}, and series expansions for strong couplings $g\gg 1$~\cite{strong_coupling_On}.
This behaviour  translates into exponentially-decaying spin-spin correlations and an energy gap in the spectrum, regardless of the value of the $g$, and thus valid for all different integer-$S$ spin chains~\cite{nlsm_heisenberg}. Conversely,  one finds a NLSM with a topological term for {\it half-integer spin}, since $\Theta=\pi$,  which modifies drastically the above properties. Building on the renormalisation-group flow to strong couplings, and thus small effective spins (as follows from $S=2/g$), Haldane conjectured that all half-integer-spin Heisenberg models should be qualitatively identical to the $S=1/2$ case,  and should thus   display algebraically-decaying  correlations and a vanishing energy gap~\cite{xxz_luttinger}. There is compelling  evidence based on different numerical methods supporting such conjecture~\cite{numerics_haldane}.

We now explore the effects of the long-range interactions on the mapping of the LRHM onto the NLSM. Let us start by reviewing the Hamiltonian approach to the NLSM described in~\cite{nlsm_hamiltonian, nlsm_heisenberg_hamiltonian}, which starts by imposing directly the constraint over the vector field by introducing two scalar fields $\alpha(x^\mu),\beta(x^\mu)$, such that $\boldsymbol{\phi}=(\sin\alpha\cos\beta,\sin\alpha\sin\beta,\cos\alpha)$. After obtaining their canonically-conjugate momenta through  $\Pi_\alpha=\partial_{\dot\alpha}\mathcal{L}$, and $\Pi_\beta=\partial_{\dot\beta}\mathcal{L}$, one finds $\mathcal{H}_{\rm NLSM}=\frac{1}{2g}\sum_\mu\left((\partial_\mu\alpha)^2+\sin^2\alpha(\partial_\mu\beta)^2\right)$ by using standard trigonometry. It is customary to introduce the angular momentum $\boldsymbol{\ell}=\boldsymbol{\phi}\times\left(\partial_\alpha\boldsymbol{\phi}\Pi_\alpha+\sin^{-2}\alpha\partial_\beta\boldsymbol{\phi}\Pi_\beta\right)$, and apply again basic trigonometric rules to obtain the final form of the NLSM Hamiltonian
\beq
\label{eq:H_NLSM}
H_{\rm NLSM}=\int{\rm d}x\frac{v}{2}\left[g\left(\boldsymbol{\ell}-\frac{\Theta}{4\pi}\partial_x\boldsymbol{\phi}\right)^2+\frac{1}{g}(\partial_x\boldsymbol{\phi})^2\right], 
\eeq
where $\boldsymbol{\phi}^2=1$, and $\boldsymbol{\ell}\cdot\boldsymbol{\phi}=0$.
Moreover, the following algebra between the vector and angular momentum fields is obtained
\beq
\begin{split}
\label{eq:phi_l_algebra}
[\phi^a(x),\phi^b(y)]&=0, \\ [\ell^a(x),\phi^b(y)]&=\ii\epsilon^{abc}\phi^c(x)\delta(x-y),\\  [\ell^a(x),\ell^b(y)]&=\ii\epsilon^{abc}\ell^c(x)\delta(x-y), 
\end{split}
\eeq
which follows from the canonical commutation relations of the scalar fields and their conjugate momenta. 

The goal now is to find a particular NLSM Hamiltonian~\eqref{eq:H_NLSM} starting from the microscopic LRHM~\eqref{eq:LRHM}, and introducing  the following  spin operators 
\beq
\label{eq:l_s_ops}
\boldsymbol{\ell}_i=\frac{1}{2a}\left(\boldsymbol{S}_{2i+1}+\boldsymbol{S}_{2i}\right),\hspace{2ex} \boldsymbol{\phi}_i=\frac{1}{2S}\left(\boldsymbol{S}_{2i+1}-\boldsymbol{S}_{2i}\right),
\eeq
which represent small and rapid fluctuations of the local spin density, and slow fluctuations of the staggered spin density, respectively. Using the $\frak{su}(2)$ spin algebra algebra, one finds that these operators satisfy the constraints
\beq
\label{eq:constraints_l_phi_lattice}
\boldsymbol{\ell}_i\cdot\boldsymbol{\phi}_i=0,\hspace{2ex}\boldsymbol{\phi}_i^2=1+\frac{1}{S}-\frac{\boldsymbol{\ell}_i^2}{S^2},
\eeq
which coincide with those of Eq.~\eqref{eq:H_NLSM} in the large-$S$ and  continuum $a\to 0$ limits. Moreover, the correct algebra is also recovered in these  limits, since 
\beq
\begin{split}
[\phi^a_i,\phi^b_j]&=\ii\epsilon^{abc}\frac{\ell^c_i}{S^2}\frac{\delta_{i,j}}{2a},\\ 
[\ell^a_i,\phi^b_j]&=\ii\epsilon^{abc}\phi^c_i\frac{\delta_{i,j}}{2a}, \\ 
 [\ell^a_i,\ell^b_j]&=\ii\epsilon^{abc}\ell^c_i\frac{\delta_{i,j}}{2a},
\end{split}
\eeq
lead to Eqs.~\eqref{eq:phi_l_algebra} in the above limits, where $\frac{\delta_{i,j}}{2a}\to \delta(x-y)$. It is then clear that the local and staggered magnetisation~\eqref{eq:l_s_ops} shall play a key role in the mapping of the LRHM onto the NLSM.

Let us rewrite the LRHM~\eqref{eq:LRHM} as $H_{\rm LRHM}=\sum_{i}\sum_r \tilde{J}_{r}\boldsymbol{S}_i\cdot\boldsymbol{S}_{i+r},$ which assumes that the spin-spin couplings are translationally invariant, and will thus describe the physics of the bulk of a large trapped-ion chain where such an approximation becomes valid (see the section below). Using the spin operators~\eqref{eq:l_s_ops}, and partitioning the sum into even- and odd-spaced spin pairs, we find
\begin{widetext}
\beq
\begin{split}
H_{\rm LRHM}&=\sum_{i,r}\tilde{J}_{2r+1}\left(a^2\boldsymbol{\ell}_{i}\cdot \boldsymbol{\ell}_{i+r\phantom{+1}}+Sa\big(\boldsymbol{\ell}_{i}\cdot \boldsymbol{\phi}_{i+r\phantom{+1}}-\boldsymbol{\phi}_{i}\cdot \boldsymbol{\ell}_{i+r\phantom{+1}}\big)-S^2\boldsymbol{\phi}_{i}\cdot \boldsymbol{\phi}_{i+r\phantom{+1}}\right)\\
&+\sum_{i,r}\tilde{J}_{2r+1}\left(a^2\boldsymbol{\ell}_{i}\cdot \boldsymbol{\ell}_{i+r+1}-Sa\big(\boldsymbol{\ell}_{i}\cdot \boldsymbol{\phi}_{i+r+1}-\boldsymbol{\phi}_{i}\cdot \boldsymbol{\ell}_{i+r+1}\big)-S^2\boldsymbol{\phi}_{i}\cdot \boldsymbol{\phi}_{i+r+1}\right)\\
&+\sum_{i,r}\tilde{J}_{2r+2}\left(a^2\boldsymbol{\ell}_{i}\cdot \boldsymbol{\ell}_{i+r+1}-Sa\big(\boldsymbol{\ell}_{i}\cdot \boldsymbol{\phi}_{i+r+1}+\boldsymbol{\phi}_{i}\cdot \boldsymbol{\ell}_{i+r+1}\big)+S^2\boldsymbol{\phi}_{i}\cdot \boldsymbol{\phi}_{i+r+1}\right)\\
&+\sum_{i,r}\tilde{J}_{2r+2}\left(a^2\boldsymbol{\ell}_{i}\cdot \boldsymbol{\ell}_{i+r+1}+Sa\big(\boldsymbol{\ell}_{i}\cdot \boldsymbol{\phi}_{i+r+1}+\boldsymbol{\phi}_{i}\cdot \boldsymbol{\ell}_{i+r+1}\big)+S^2\boldsymbol{\phi}_{i}\cdot \boldsymbol{\phi}_{i+r+1}\right).
\end{split}
\eeq
\end{widetext}
The next step is to take the continuum limit, and make a gradient expansion keeping terms that are $\mathcal{O}(a^2)$. Then, it suffices to consider the following approximations for different pairs  of integers $\{n,m\}$, namely
\begin{widetext}
\beq
\begin{split}
&\boldsymbol{\ell}_{n}\cdot \boldsymbol{\ell}_{n+m\hspace{1ex}}\to \boldsymbol{\ell}(x)\cdot \boldsymbol{\ell}(x+(2a)m)\hspace{1ex}\approx \boldsymbol{\ell}^2(x),\\
&\boldsymbol{\phi}_{n}\cdot \boldsymbol{\phi}_{n+m}\to \boldsymbol{\phi}(x)\cdot \boldsymbol{\phi}(x+(2a)m)\approx \boldsymbol{\phi}^2(x)+(2a)m\boldsymbol{\phi}(x)\cdot \partial_x\boldsymbol{\phi}(x)+\half (2a)^2m^2\boldsymbol{\phi}(x)\cdot \partial^2_x\boldsymbol{\phi}(x)\approx\left(1+\frac{1}{S}-\frac{\boldsymbol{\ell}^2(x)}{S^2}\right)-2a^2m^2(\partial_x\boldsymbol{\phi})^2, \\
&\boldsymbol{\phi}_{n}\cdot \boldsymbol{\ell}_{n+m\hspace{0.25ex}}\to \boldsymbol{\phi}(x)\cdot \boldsymbol{\ell}(x+(2a)m)\hspace{0.5ex}\approx\boldsymbol{\phi}(x)\cdot \boldsymbol{\ell}(x)+(2a)m\boldsymbol{\phi}(x)\cdot \partial_x\boldsymbol{\ell}(x)=-(2a)m\boldsymbol{\ell}(x)\cdot \partial_x\boldsymbol{\phi}(x),\\
&\boldsymbol{\ell}_{n}\cdot \boldsymbol{\phi}_{n+m\hspace{0.25ex}}\to \boldsymbol{\ell}(x)\cdot \boldsymbol{\phi}(x+(2a)m)\hspace{0.5ex}\approx\boldsymbol{\ell}(x)\cdot \boldsymbol{\phi}(x)+(2a)m\boldsymbol{\ell}(x)\cdot \partial_x\boldsymbol{\phi}(x)=(2a)m\boldsymbol{\ell}(x)\cdot \partial_x\boldsymbol{\phi}(x).
\end{split}
\eeq
\end{widetext}
To arrive at these expressions, we have performed the corresponding Taylor expansions for $a\to 0$, used the lattice constraints~\eqref{eq:constraints_l_phi_lattice}, and considered integration by parts under  $\sum_i (2a)\to\int{\rm d}x$ assuming that the fields vanish at the boundaries of the sample. Under these approximations, the Hamiltonian of the LRHM becomes
\begin{widetext}
\beq
H_{\rm LRHM}\approx\int{\rm d}x\left[2a\sum_r \tilde{J}_{2r+1}\boldsymbol{\ell}^2(x)-Sa\sum_r\tilde{J}_{2r+1}\big(\boldsymbol{\ell}(x)\cdot\partial_x\boldsymbol{\phi}(x)+\partial_x\boldsymbol{\phi}(x)\cdot \boldsymbol{\ell}(x)\big)-\frac{Sa^2}{2}\sum_r(-1)^r r^2J_{r}(\partial_x\boldsymbol{\phi}(x))^2\right].
\eeq
\end{widetext}
By direct comparison with the Hamiltonian~\eqref{eq:H_NLSM} of the NLSM, we find  a system of three algebraic equations that leads to the following NLSM parameters
\beq
\begin{split}
\label{eq:NLSM_parameters}
v&=2aS\sqrt{\sum_{r\hspace{0.2ex}{\rm odd}} \tilde{J}_{r}\left(\sum_{r\hspace{0.2ex}{\rm odd}}  r^2\tilde{J}_{r}-\sum_{r\hspace{0.2ex}{\rm even}}  r^2\tilde{J}_{r}\right)},\\
 g&=\frac{2}{S}\sqrt{\sum_{r\hspace{0.2ex}{\rm odd}} \tilde{J}_{r}\left(\sum_{r\hspace{0.2ex}{\rm odd}}  r^2\tilde{J}_{r}-\sum_{r\hspace{0.2ex}{\rm even}}  r^2\tilde{J}_{r}\right)^{-1}}, \\
 \Theta&=2\pi S.
\end{split}
\eeq
As a consistency check, we note that in the nearest-neighbor limit $\tilde{J}_r=\tilde{J}\delta_{r,1}$ with $\tilde{J}>0$, we recover the same parameters for the mapping of the Heisenberg model onto the NLSM, namely $v=2\tilde{J}a S, g=2/S$, and $\Theta=2\pi S$, which lead to Haldane's conjecture.

In the main text, we evaluate  the above coupling constant $g$  for the particular spin-spin couplings that arise in the trapped-ion scenario. It is also very interesting to consider
long-range interactions that decay with a power-law of the distance
\beq
\tilde{J}_r = \frac{ \tilde{J}_1}{r^s}, 
\label{2}
\eeq
with an exponent $s>0$. In the thermodynamic limit,  the  series appearing in Eq.(\ref{eq:NLSM_parameters}) can be expressed in terms of Dirichlet  $\eta$ and $\lambda$ functions,
that in turn are related in a simple way to the Riemann zeta function $\zeta$, 
\begin{widetext}
\barray 
 \sum_{r {\rm odd}} \tilde{J}_n  & = & \sum_{n=0}^\infty  \frac{ \tilde{J}_1}{(2 n + 1)^s} = \tilde{J}_1 \,  \lambda(s) =  \tilde{J}_1 \,   (1 - 2^{-s}) \zeta(s) , \quad {\rm Re}\,  s > 1 
 \label{3} \\
 \sum_{r {\rm odd}} r^2 \tilde{J}_r -   \sum_{r {\rm even}} r^2 \tilde{J}_r  & = & 
\sum_{n=1}^\infty  \frac{ \tilde{J}_1 (-1)^{1+n}}{n^{s-2}}  = \tilde{J}_1  \, \eta(s-2) =  \tilde{J}_1  (1- 2^{3-s}) \zeta(s-2) , \quad {\rm Re} \, s > 2 
\label{4} 
 \earray 
 \end{widetext}
The convergence of these two series requires that $s >2$.  In the  limit $s \rightarrow 2$, the series
Eq.(\ref{4})  is Abel convergent, which means that it can be regularized by adding  a term $e^{- x n}$, with $x>0$,  and then taking  the limit
$x \rightarrow 0$. This gives the well known result $1 - 1 + 1+ \dots = 1/2$. In this case one obtains the 
the Haldane-Shastry model where the exchange couplings decreases as in inverse square distance
\cite{Haldane_shastry}. The values of $v$ and $g$ are given by 
\beq
s=2 \Longrightarrow v = 4 \pi a J_1, \qquad g = 2 \pi.
\label{5}
\eeq

\section{Distance decay of the effective spin-spin interactions}
\label{app:C}

In this Appendix, we derive explicit formulas for the distance dependence of the phonon-mediated spin-spin interactions in the trapped-ion crystal~\eqref{eq:spin-spin_int}. 
If we consider the small Lamb-Dicke parameter, which is particularly the  case for heavy ions such as $^{171}$Yb$^+$, and also take into account additional off-resonant terms in the MS scheme~\cite{varenna_proceedings}, the spin-spin interactions can be written as
\beq
\label{eq:J_full}
J_{ij}=\frac{|\Omega_{\rm L}|^2}{2}\omega_{\rm R}\sum_n\frac{\mathcal{M}_{i,n}\mathcal{M}_{j,n}^*}{\mu^2-\omega_{n,x}^2}+{\rm c.c.},
\eeq
where we have introduced the recoil energy $\omega_{\rm R}=(\Delta{\bf k}\cdot {\bf e}_x)^2/2m$, and the symmetric beatnote  of the MS beams $\omega_{\rm b}=\omega_0+\mu$,$\omega_{\rm r}=\omega_0-\mu$, which corresponds to  $\mu=\omega_{n,x}-\delta_n$  according to our previous notation. It is straightforward to see that in the resolved-sideband limit considered throughout this work $|\delta_n|\ll\omega_n$, one can approximate $\mu^2-\omega_{n,x}^2=(\mu+\omega_{n,x})(\mu-\omega_{n,x})\approx -2\omega_{n,x}\delta_n$. Hence,  by using the expression of the MS forces~\eqref{eq:forces} for small Lamb-Dicke parameters $\eta_n\ll 1$, we see that Eq.~\eqref{eq:J_full}  is equivalent  to the spin-spin couplings derived in  Eq.~\eqref{eq:spin-spin_int}. Nonetheless,  it will  be useful to use the full expression~\eqref{eq:J_full} instead of~\eqref{eq:spin-spin_int} in the  derivation of  the distance decay of the spin-spin couplings. Let us emphasise that, although a power-law decay with a tunable exponent $s\in[0,3]$, namely $J_{ij}=\tilde{J}_0/|i-j|^s$, serves as a convenient approximation in experiments~\cite{Ising_exp_penning, xy_dynamics, ions_spectroscopy,  variable_range}, special care must be taken when such expressions are to be extrapolated to the thermodynamic limit in theoretical studies. This is particularly so for the evaluation of the NLSM parameters~\eqref{eq:NLSM_parameters}, which crucially depend on the long-range tail of the spin-spin couplings. 

In Ref.~\cite{J_distance_continuum_limit}, it was shown that if  Eq.~\eqref{eq:spin-spin_int} is approximated further by considering  $\mu^2-\omega_{n,x}^2\approx -2\omega_{n,x}\delta_n\approx-2\omega_{x}\delta_n$, it is possible to derive an analytical estimate of the spin-spin couplings by using a continuum limit, and substituting the sum over the normal modes by an integral that can be evaluated by an extension to the complex plane. Provided that $\mu<\omega_{n,x}$, it was shown that $J_{ij}$ has two contributions: a term that displays a dipolar decay, and another one that shows an exponential tail with a characteristic decay length dominated by the detuning of the spin-dependent force with respect to the lowest-energy zig-zag mode. It is by varying this characteristic  length  that the spin-spin couplings show a variable range that can be seen as  an effective power law $J_{ij}=\tilde{J}_0/|i-j|^\alpha$ that is slower than the dipolar decay $\alpha<3$ for sufficiently small chains. However, let us remark again that for theoretical extrapolations to very large ion chains, one should use directly the correct distance dependence.

We will now show that a similar result can be obtained without making the  approximation $\mu^2-\omega_{n,x}^2\approx-2\omega_{x}\delta_n$ in Eq.~\eqref{eq:J_full}, and independently of the choice of $\mu\lessgtr\omega_{n,x}$, as far as the force is far from the resonance with any  mode in the vibrational band $\mu\neq\omega_{n,x}$. We shall not resort to a continuum limit, but partially resume the most relevant terms of Eq.~\eqref{eq:J_full}.
Since we are interested in the predictions of the mapping of the spin chain onto the NLSM~\eqref{eq:NLSM_parameters}, which are only valid for the bulk of the trapped-ion crystal, we use a homogeneous lattice spacing $a_0$ that corresponds to the  distance between two neighbouring ions in the centre of the chain.  Following~\cite{J_distance_continuum_limit}, we describe the normal-mode  displacements and frequencies of the ion crystal by
\beq
\label{eq:normal__modes}
\mathcal{M}_{j,n}=\frac{1}{\sqrt{N}}\ee^{\ii qa_0j},\hspace{3ex}\omega_{n,x}=\sqrt{\tilde{\omega}_x^2+2\beta_x\omega_x^2\sum_{d=1}^{N/2}\frac{c_d}{d^3}\cos(qa_0d)}, 
\eeq
where we have used periodic boundary conditions, such that it is possible to introduce   the quasi-momentum within the Brillouin zone  $q=2\pi n/ Na_0\in{\rm BZ}=[0,2\pi/a_0)$, the coefficients  $c_d=(1-\delta_{d,N/2})+\delta_{d,N/2}/2$, and a renormalised trap frequency $\tilde{\omega}_x=\omega_x(1-2\beta_x\sum_{d}c_d/d^3)^{1/2}$ that depends on the stiffness parameter $\beta_x=(e^2/4\pi\epsilon_0a_0)/m\omega_x^2a_0^2$~\cite{porras_spin_models_ions}. We  now substitute Eq.~\eqref{eq:normal__modes} in the expression~\eqref{eq:J_full}, and use the geometric Taylor series  for $|\lambda|<1$, in order  to express the spin-spin couplings as follows
\beq
J_{i,j}=\frac{J_0}{N}\sum_{q\in{\rm BZ}}\sum_{n=0}^{\infty}\ee^{\ii qa_0(i-j)}\lambda^n\left(\sum_{d=1}^{N/2}\frac{c_d}{d^3}\cos(qa_0d)\right)^n+{\rm c.c.},
\eeq
where  we have introduced two important  parameters in our calculations
\beq
\label{eq:parameters}
J_0=\frac{|\Omega_{\rm L}|^2\omega_{\rm R}}{2(\mu^2-\tilde{\omega}_{x}^2)}, \hspace{3ex}
\lambda=\frac{2\beta_x\omega_x^2}{(\mu^2-\tilde{\omega}_{x}^2)}.
\eeq
If $\mu^2>\tilde{\omega}_{x}^2+2\beta_x\omega_x^2$, the beatnote of the MS laser beams will be off-resonant with respect to the whole vibrational branch.  Therefore, assuming that $|\lambda|<1$ requires working at sufficiently-large MS detunings $|\delta_n|>\beta_x\omega_x$, a fact that is in any case required to minimise the error of the quantum  simulator for a spin chain~\cite{porras_spin_models_ions}.

\begin{figure*}
\centering
\includegraphics[width=1.9\columnwidth]{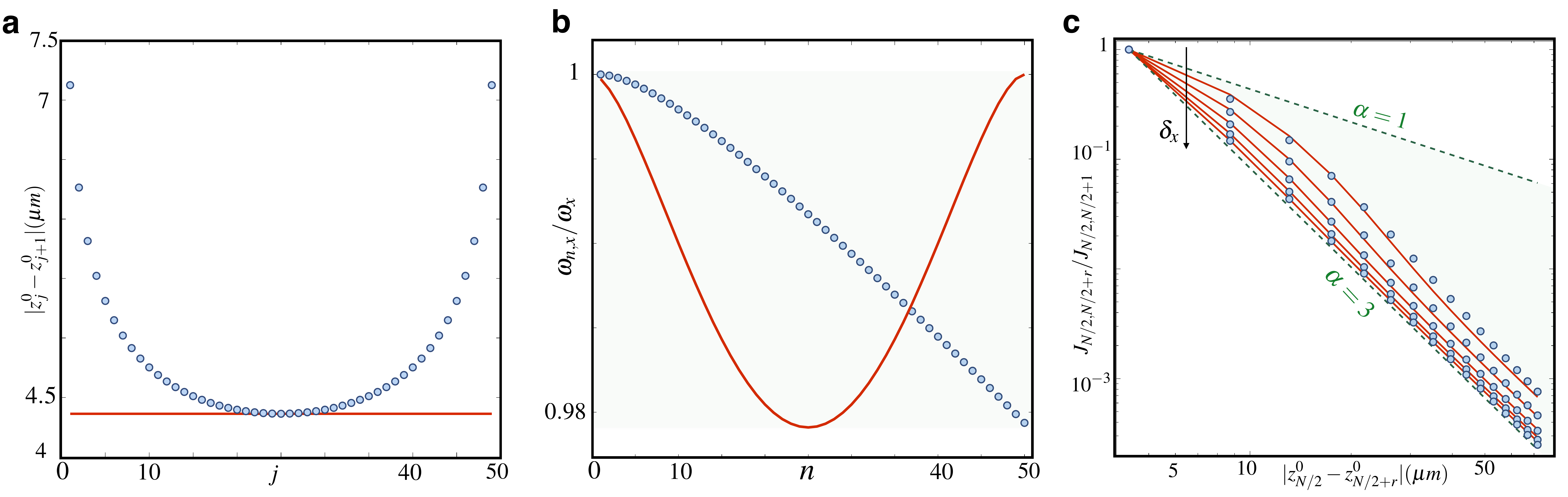}
\caption{ {\bf Distance dependence of the spin-spin couplings:} {\bf (a)} Distance between neighbouring ions $|z_j^0-z_{j+1}^0|$ as a function of the lattice index $j\in\{1,\cdots,49\}$ for a chain of $N=50$  Yb ions. The blue circles are obtained from the exact equilibrium positions  $z_j^0=\ell_z u_j$ by solving  $u_j-\sum_{k\neq j}(u_j-u_k)/|u_j-u_k|^3=0$, where $\ell_z=(e^2/4\pi\epsilon_0m\omega_z^2)^{1/3}$. The red line stands for the theoretical model of a homogeneous ion chain $z_j^0=a_0 j$ with constant lattice spacing given by $a_0={\rm min}\{|z_j^0-z_{j+1}^0|\}$. {\bf (b)} Normalized vibrational frequencies $\omega_{n,x}/\omega_x$ as a function of the normal-mode index $n\in\{1,\cdots, 50\}$. The blue circles represent the exact normal modes obtained by solving $\sum_{ij}\mathcal{M}_{i,n}\mathbb{K}_{i,j}\mathcal{M}_{j,m}=\omega_{n,x}^2\delta_{n,m}$, where $\mathbb{K}_{i,j}/\omega_x^2=(1-\delta_{i,j})\beta_x/|z_i^0-z_j^0|^3+\delta_{i,j}(1-\sum_{l\neq i}\beta_x/|z_i^0-z_l^0|^3)$ is obtained from the numerical solution of the  equilibrium positions in the inhomogeneous crystal. The red line stands for the theoretical model for a homogeneous periodic chain with the normal modes given in Eq.~\eqref{eq:normal__modes}. {\bf (c)} Normalised spin-spin couplings $J_{i, j}/J_{i,i+1}$ between the central ion $i=N/2=25$ and its bulk neighbours  $j=N/2+r$, with $r\in\{1,\cdots 16\}$,  as a function of their respective distance $|z_i^0-z_{j}^0|$, and for different values of the MS detuning $\delta_x=\omega_x-\mu$ with respect to the center-of-mass mode, $\delta_x/2\pi\in\{62.5, 125, 250, 500, 1000\}$kHz, increasing in the direction of the arrow. The blue circles are obtained by solving Eq.~\eqref{eq:J_full} using the previous normal modes and frequencies. The red lines stand for our analytical result~\eqref{eq:J_distance} without any fitting parameter, but rather using the microscopic values for Eqs.~\eqref{eq:parameters} and~\eqref{eq:decay_length}. The green dashed lines correspond to a power-law decay $J_r\propto 1/r^s$ for two different exponents $s=3$, and $s=1$.    }
\label{fig_sm_2}
\end{figure*}

To proceed further, we approximate$\left(\sum_d\frac{c_d}{d^3}\cos(qa_0d)\right)^n\approx \cos^n(qa_0)+n\sum_{d>1}\frac{c_d}{d^3}\cos(qa_0d)$, which is justified given the fast dipolar decay of these couplings. By finally making use of the binomial theorem, we find that $J_{ij}=J_{ij}^{(1)}+J_{ij}^{(2)}$, where
\begin{widetext}
\beq
\label{eq:couplings}
\begin{split}
J^{(1)}_{ij}&=\frac{J_0}{N}\sum_{q\in{\rm BZ}}\sum_{n=0}^{\infty}\sum_{k=0}^{n}\left(\frac{\lambda}{2}\right)^n\left(\begin{array}{c}n \\k\end{array}\right)\ee^{\ii qa_0\big((i-j)+n-2k\big)}+{\rm c.c.},\\
J^{(2)}_{ij}&=\frac{J_0}{N}\sum_{q\in{\rm BZ}}\sum_{n=0}^{\infty}\sum_{k=0}^{n-1}\sum_{d=2}^{N/2}\frac{c_d}{d^3}n\left(\frac{\lambda}{2}\right)^n\left(\begin{array}{c}n-1 \\k\end{array}\right)\left(\ee^{\ii qa_0\big((i-j)+n-1-2k+d\big)}+\ee^{\ii qa_0\big((i-j)+n-1-2k-d\big)}\right)+{\rm c.c.}.\\
\end{split}
\eeq
\end{widetext}
To evaluate these expressions, we need to make use of the sum of angles in the complex unit circle $\sum_{q\in{\rm BZ}}\ee^{\ii qa_0x}=N\delta_{x,0}$, and a number of combinatorial identities. By  introducing $r=i-j$ and  $J_{ij}=J_{i-j}=:J_{r}$, we find $J_r=J_{r}^{(1)}+J_{r}^{(2)}$ with
\beq
\label{eq:J_distance}
J^{(1)}_{r}= 2|J_0|({\rm sgn}\big(\lambda)\big)^{1+|r|}\ee^{-\frac{|r|a_0}{\xi_0}},
\eeq
and
\begin{widetext}
\beq
J^{(2)}_{r}= \sum_{\delta r= 2-r}^{N/2-r}\left|J_0\lambda\right|({\rm sgn}\big(\lambda)\big)^{1+|\delta r|}\frac{c_{r+\delta r}}{|r+\delta r|^3}\left(\frac{\sqrt{1-\lambda^2}+|\delta r|(1-\lambda^2)}{(1-\lambda^2)^2}\right)\ee^{-\frac{|\delta r|a_0}{\xi_0}}\theta(r-2),
\eeq
\end{widetext}
where  we have introduced the following decay length associated to the exponential terms
\beq
\label{eq:decay_length}
\xi_0=-\frac{a_0}{\log\left(\frac{1-\sqrt{1-\lambda^2}}{|\lambda|}\right)},
\eeq
and  the Heaviside step function, $\theta(x)=1$ if $x\geq0$ and zero elsewhere.

Let us now comment on the different possible regimes. In the limit of $\lambda\to 0$, the exponential terms decay very rapidly since $\xi_0\to0$, and we find $J^{(1)}_{r}\approx |J_0\lambda|\delta _{r,1}$,  $J^{(2)}_{r}\approx |J_0\lambda|/|r|^3\theta(r-2)$. Therefore,  for very large detunings of the MS force such that  $\lambda\to 0$, we recover the well-known antiferromagnetic dipolar limit of the phonon-mediated spin-spin interactions $J_r\approx \tilde{J}_0/|r|^3$, where $\tilde{J}_0=|J_0\lambda|=|\Omega_{\rm L}|^2\omega_{\rm R}\beta_x\omega_x/8\omega_x\delta_x^2>0$~\cite{porras_spin_models_ions}. On the other hand, for finite but still small $|\lambda|\ll1$, we find that the leading order corrections to the dipolar tail come from an exponentially-decaying term
\beq
J_r\approx\frac{|J_0\lambda|}{|r|^3}+2|J_0|\big({\rm sgn}(\lambda)\big)^{1+|r|}\ee^{-\frac{|r|a_0}{\xi_0}}\theta(r-2),
\eeq
a result similar to that found in~\cite{J_distance_continuum_limit}, but valid for red/blue detunings without making any further approximation to Eq.~\eqref{eq:J_full}. The detuning of the MS forces can be understood as an effective mass $m_{\rm eff}$ of the phonons that carry the spin-spin interactions. This effective mass would naturally account for an exponential decay  of the interactions with length $\xi\propto 1/m_{\rm eff}$, which would be combined with the natural dipolar decay associated to a system for transversally-oscillating charges (i.e. effective dipoles). As $|\lambda|$ grows larger $0<|\lambda|<1$, the complete expression in Eq.~\eqref{eq:J_distance} must be considered, as the different exponentials can lead to considerable modifications of the spin-spin couplings.

We should now test the validity of our result~\eqref{eq:J_distance}  by comparing with a numerically-exact evaluation of the spin-spin couplings~\eqref{eq:J_full} using the equilibrium positions and normal modes~\cite{normal_modes}  for  a  ion crystal with the realistic parameters introduced in the main text. We  consider $N=50$ $^{171}$Yb$^+$ ions in a linear Paul trap with frequencies $\omega_z/2\pi=0.1$MHz, and $\omega_x/2\pi=5$MHz, which form an inhomogeneous chain with minimal lattice spacing $a_0=4.4\mu$m corresponding to two neighbouring ions in the centre of the chain (see Fig.~\ref{fig_sm_2}{\bf (a)}). In this figure, we see that the lattice spacing of the finite ion chain is inhomogeneous, and varies considerably when approaching the chain boundaries. We also display the theoretical model of a constant lattice spacing to describe the bulk of the ion crystal. The  vibrational frequencies $\omega_{n,x}$ are displayed in Fig.~\ref{fig_sm_2}{\bf (b)}, where we compare the exact numerical values with those  obtained by using a periodic ion chain with homogeneous lattice spacing~\eqref{eq:normal__modes}. We observe that the vibrational bands have the same width in both cases, while the doubling of the vibrational frequencies is a consequence of the periodic boundary conditions as opposed to the open boundary conditions of a realistic chain. Despite the clear differences in Figs.~\ref{fig_sm_2}{\bf (a)} and~\ref{fig_sm_2}{\bf (b)}, we  shall now show that the theoretical model gives very accurate results for the spin-spin interactions of bulk ions in the inhomogeneous ion chain. In Fig.~\ref{fig_sm_2}{\bf (c)}, we compare the exact spin-spin couplings for the inhomogeneous ion chain~\eqref{eq:J_full}, with the analytical estimates~\eqref{eq:J_distance}   based on the theoretical model of the periodic homogeneous ion chain~\eqref{eq:normal__modes}. The agreement between both values is quite remarkable, given the relatively small size of the ion chain, and the clear differences displayed in Figs.~\ref{fig_sm_2}{\bf (a)} and~\ref{fig_sm_2}{\bf (b)}. Let us highlight that the considered detunings in the MS forces correspond to   $\lambda\in[0.04,0.4]$, and thus to some instances where the parameter $\lambda$ is far from being a small perturbation.

We can also identify the qualitative behaviour described below Eq.~\eqref{eq:J_distance}: for very large MS detunings, the distance-dependence can be reliably approximated by a dipolar law. As the detunings decrease, and thus the relevant parameter $\lambda$ increases, the contribution of an exponential tail to the spin-spin couplings starts playing a role. This becomes apparent for the couplings $J_r$ at small distances $r a_0\ll\xi_0$, where we see a decay that is much slower than the dipolar law. However, at large distances $r a_0\gg\xi_0$, the contribution from the exponential tail is suppressed, and one recovers the dipolar power law. These numerical results confirm that the analytical estimates~\eqref{eq:J_distance} are more accurate than a fit to a power-law decay $J_r\propto 1/r^s$ with a varying exponent $s\in[0,3]$. Let us finally remark that, in order to obtain an analytical expression for even larger interaction ranges, one should take into account further terms in the approximation above Eq.~\eqref{eq:couplings}, which may become relevant for sufficiently-small  detunings. In any case, such small MS detunings compromise the validity of a pure effective spin model, as  errors due to a thermal phonon population start playing a dangerous role~\cite{porras_spin_models_ions}.



\begin{references}
\bibitem{comment_staggered}
For $N/2$ odd, we require a staggered version of the transverse field $\sum_ih\sigma_i^y\to \sum_ih(-1)^i\sigma_i^y$. Experimentally, it may be easier to  substitute $\sigma_j^y\to\sigma_j^z$, which could be implemented through local ac-Stark shifts in the trapped-ion experiment, and would require an initial state  $\ket{\psi_0}=\ket{\downarrow\uparrow\downarrow\cdots}$.
\bibitem{comment_odd} For $N$ odd, the groundstate can display additional degeneracies, as can be easily seen from the Jordan-Wigner fermion solution of the nearest-neighbor XY model $\phi_{\rm s}=\pi/2$~\cite{lsm},  when applied to open boundary conditions. We avoid these by considering only $N$ even.
\bibitem{comment_c_nn}
We have checked that our numerical routine yields the predicted central charge  $c=1$ for the nearest-neighbour limit of the Heisenberg model, using the same bond dimension $\chi=256$ of the Matrix-Product-State.
\end{references}
\end{document}